\documentclass[a4paper,11pt]{article}
\pdfoutput=1 

\usepackage{jheppub} 
\usepackage{amsmath}
\allowdisplaybreaks[4]        
\usepackage{amssymb}
\usepackage{euscript}     
\usepackage{color}         
\usepackage{tensor}        
\usepackage{amsthm}
\usepackage{bbm}
\usepackage{float}
\usepackage{pgfplots}
\usepackage{graphicx,subcaption}
\usepackage{mathtools}	
\usepackage{subcaption}
\usepackage{hyperref}
\hypersetup{colorlinks=true,allcolors=blue}

\usepackage[most]{tcolorbox}
\tcbuselibrary{skins, breakable, theorems}

\usepackage[header,title,page,titletoc]{appendix}  
\usepackage[T1]{fontenc} 
\usepackage[numbers]{natbib}  
\newcommand{\imineq}[2]{\vcenter{\hbox{\includegraphics[height=#2ex]{#1}}}}

\definecolor{browna}{rgb}{0.76,0.72,0.65}
\definecolor{brownb}{rgb}{0.71,0.69,0.65}
\definecolor{SpringGreen}{rgb}{0.95,0.97,0.95}
\definecolor{OliverGreen}{rgb}{0.09,0.34,0.09}
\definecolor{LeftGreen}{rgb}{0.13,0.54,0.13}

\newtcbtheorem[no counter]{theorem}{Theorem}{
	enhanced,
	sharp corners,
	attach boxed title to top left={
		yshifttext=-1mm
	},
	colback=white,
	colframe=browna,
	fonttitle=\bfseries,
	boxed title style={
		sharp corners,
		size=small,
		colback=browna,
		colframe=browna,
	} 
}{thm}

\newtcbtheorem[no counter]{Example}{Example}{
	enhanced,
	sharp corners,
	attach boxed title to top left={
		yshifttext=-1mm
	},
	colback=white,
	colframe=brownb,
	fonttitle=\bfseries,
	coltitle=white,
	boxed title style={
		sharp corners,
		size=small,
		colback=brownb,
		colframe=brownb,
	} 
}{exm}

\newtcbtheorem{myclaim}{Claim -- }
{    coltitle=OliverGreen,    
	colback=SpringGreen,    colframe=LeftGreen,    detach title,    boxrule=0pt,    leftrule=2pt,    attach title to upper,    sharp corners,    left=1mm,}
{claim}

\DeclareMathOperator{\Tr}{Tr}

\newcommand{\ri}{\mathrm{i}}
\renewcommand{\th}{\theta}

\newcommand{\cob}{\delta}

\newcommand{\vep}{\varepsilon}

\newcommand{\hf}{\frac{1}{2}}

\newcommand{\del}{\partial}

\newcommand{\bra}{\langle}
\newcommand{\ket}{\rangle}
\newcommand{\la}{\lambda}

\newcommand{\h}[1]{\widehat{#1}}
\newcommand{\bt}{\beta}

\newcommand{\rt}[1]{\sqrt{#1}}
\newcommand{\cO}{\mathcal{O}}
\newcommand{\cZ}{\mathcal{Z}}

\newcommand{\cB}{\mathcal{B}}

\newcommand{\cN}{\mathcal{N}}

\newcommand{\qb}[1]{\Bigl[~\begin{matrix}#1\end{matrix}~\Bigr]_q}

\newcommand{\id}{\mathbbm{1}}

\makeatletter
\gdef\@fpheader{}
\makeatother
\begin{document}
\begin{flushright}
YITP-25-71,~RIKEN-iTHEMS-Report-25
\\
\end{flushright}

\title{Finite $N$ Bulk Hilbert Space in ETH Matrix Model for double-scaled SYK:
\vskip 0.1in
\Large{\it Null States, State-dependence and Krylov State Complexity}} 

\author{Masamichi Miyaji$^{1,2}$, Soichiro Mori$^3$ and Kazumi Okuyama$^4$}

\affiliation{$^1$Yukawa Institute for Theoretical Physics, Kyoto University, Kyoto, 606-8267, Japan}
\affiliation{$^2$ RIKEN Center for Interdisciplinary Theoretical and Mathematical Sciences (iTHEMS),
RIKEN, 2-1 Hirosawa, Wako, Saitama, 351-0198, Japan}
\affiliation{$^3$Department of Physics, Nagoya University,
Nagoya, Aichi 464-8602, Japan}
\affiliation{$^4$Department of Physics, 
Shinshu University, 3-1-1 Asahi, Matsumoto 390-8621, Japan}
\emailAdd{masamichi.miyaji@gmail.com, soichiro@eken.phys.nagoya-u.ac.jp, kazumi@azusa.shinshu-u.ac.jp}

\abstract{We extend the notion of chord number in the strict large $N$ double-scaled Sachdev-Ye-Kitaev (DSSYK) model to the corresponding finite $N$ ETH matrix model. The chord number in the strict large $N$ DSSYK model is known to correspond to the discrete length of the Einstein-Rosen bridge in the gravity dual, which reduces to the renormalized geodesic length in JT gravity at weak coupling. At finite $N$, these chord number states form an over-complete basis of the non-perturbative Hilbert space, as the structure of the inner product gets significantly modified due to the Cayley-Hamilton theorem: There are infinitely many null states. In this paper, by considering ``EFT for gravitational observables'' or a version of ``non-isometric code'', we construct a family of chord number operators at finite $N$. While the constructed chord number operator depends on the reference chord number state, it realizes approximate $q$-deformed oscillator algebra and reproduces semiclassical bulk geometry around the reference state. As a special case, we will show that when the reference is chosen to be the chord number zero state, the chord number operator precisely matches with the Krylov state complexity, leading to the ``ramp-slope-plateau'' behavior at late times, implying the formation of ``grey hole''.}

\maketitle


\section{Introduction}

The double-scaled SYK (DSSYK) model \cite{Berkooz:2018jqr} is the particular limit of the SYK model \cite{Sachdev:1992fk, Kitaev1, kitaev2, Maldacena:2016hyu} with $N$ Majorana fermions with $p$-body interaction. The limit is called the double-scaled limit, which takes large $N$ while the ratio $\lambda:=\frac{2p^2}{N}$ is fixed. In this limit, the ensemble average of the $k$-th moment of the Hamiltonian $\langle \Tr[H^k]\rangle$ can be expressed as the sum over chord diagrams with $q:=e^{-\lambda}$
\begin{equation}
    \langle \Tr[H^k]\rangle=\sum_{\alpha: \text{chord diagram}}q^{m_\alpha},
\end{equation}
where $m_\alpha$ is the number of self-intersections of the chord diagram $\alpha$. The chord diagrams arise from the contraction of the random coupling in the SYK. Interestingly, this calculation technique happens to have a direct interpretation of gravity. Namely, we can construct an auxiliary Hilbert space which is isomorphic to that of the large $N$ DSSYK model \footnote{More precisely, it is isomorphic to the diagonal subspace of the doubled DSSYK Hilbert space, spanned by $|E_i\rangle_L|E_i\rangle_R\in H_L\otimes H_R$.} spanned by the \emph{chord number states} $|n\rangle~(n\in\mathbb{Z}_{\geq 0})$, and the chord number corresponds to the number of lines intersecting with the ``bulk timeslice'' in the chord diagram. With this interpretation, the bulk length corresponds to the chord number, so that the boundary DSSYK Hilbert space is now spanned by those states with clear bulk interpretation. One can also add matter to the DSSYK, in which case the bulk contains additional matter chords \cite{Berkooz:2018jqr}.

The DSSYK model can be considered a regularization of the Jackiw-Teitelboim gravity whose length spectrum is continuous. Namely, the large $N$ density of the state of the DSSYK has a finite range $-E_0\leq E\leq E_0$ with $E_0=2/\sqrt{1-q}$, and the total number of states is $2^{N/2}$. The JT gravity, whose density of states is not bounded from above, has an infinite number of states. Furthermore, by taking the limit
\begin{equation}
    \text{JT gravity limit}:~q\rightarrow 1,~~E\to -E_0,
\end{equation}
the DSSYK model reduces to the JT gravity, with the chord number states becoming the geodesic length state, where the renormaliized JT gravity geodesic length $l$ is related to the chord number $n$ as $l=n\log\frac{1}{q}$ \cite{Lin:2022rbf}. Thus we can view the DSSYK model as the discretized, regularized version of the JT gravity. Alternatively, the sine-dilaton gravity with a gauge condition on the conjugate momentum operator was argued to be identical to the DSSYK model \cite{Blommaert:2024whf, Heller:2024ldz, Blommaert:2025avl}. As the density of states has a maximum, we can study the infinite temperature limit, which has some features of gravity in a de Sitter space, and there are several interesting attempts to relate the DSSYK to the gravity in de Sitter space \cite{Verlinde:2024znh}.

In the classical two-sided black hole in real-time, the volume of the maximal volume slice has been argued to be dual to a boundary quantity that measures the complexity of the state, such as quantum state complexity and the Fisher information. In the case of the DSSYK model at strict large $N$, one can explicitly identify the bulk length with the chord number, which is equal to the Krylov spread complexity with the infinite temperature thermofield double state $|\text{TFD}(0)\rangle$ as the initial state \cite{Lin:2022rbf, Rabinovici:2023yex}, displaying linear growth as expected from the classical two-sided black hole geometry. The Krylov spread complexity \cite{Balasubramanian:2022tpr, Nandy:2024htc, Baiguera:2025dkc} captures the spreading of the initial state $|\psi_0\rangle$ over the Krylov basis, which is constructed via the Gram-Schmidt orthogonalization of the ordered set of states $(H^n|\psi_0\rangle)_{n\in\mathbb{Z}_{\geq 0}}$, therefore it measures some notion of complexity of the state. This story is satisfying at strict large $N$, while the relation between the bulk geodesic length and the Krylov state complexity is unclear at finite $N$.

Let us consider the finite $N$ DSSYK model. Recent studies have uncovered that finite $N$ effects in gravity can be explained by including Euclidean wormholes to the gravitational path-integral \cite{Saad:2018bqo, Saad:2019lba, Penington:2019kki, Almheiri:2019qdq, Saad:2019pqd, Balasubramanian:2022gmo}, so that the late time ramp of the spectral form factor and the entropy of the Hawking radiation were explained via gravity. Let us consider DSSYK with large $N$ but keeping $N$ finite. It turns out that multi-trace observables in the DSSYK model have large fluctuation compared to the matrix models \cite{Berkooz:2020fvm} suppressed by polynomial in $N$ instead of exponential suppression as in random matrix theory. Such large fluctuation cannot be explained via usual Euclidean wormholes. Thus instead of considering the original DSSYK model, we will consider the ETH matrix model for the DSSYK model \cite{Jafferis:2022wez, Jafferis:2022uhu}. The ETH matrix model reproduces the DSSYK at the strict large $N$ limit and is described by the two-matrix model at finite $N$. As a matrix model, the multi-trace correlation is exponentially suppressed in $N$, which is clearly different from the original DSSYK model. Note that for the emergence of the bulk at strict large $N$, we only need the chord diagram structure and related algebra, but not the full DSSYK. In the ETH matrix model for the DSSYK, various geometric quantities also become discrete, such as the length of the closed geodesic in the double trumpet \cite{Okuyama:2023byh} as well as the Weil-Petersson volume \cite{Okuyama:2023kdo}. We will review them in section \ref{section:ETHmatrixmodel}.

This paper investigates the chord number operator (geodesic length operator) in the ETH matrix model for the DSSYK at finite $N$, and studies the relation to the Krylov state complexity \cite{Caputa:2021sib, Balasubramanian:2022tpr,Rabinovici:2022beu, Lin:2022rbf, Bhattacharjee:2022ave,Balasubramanian:2022dnj, Erdmenger:2023wjg, Hashimoto:2023swv, Caputa:2023vyr,  Rabinovici:2023yex, Nandy:2024htc, Camargo:2024deu, Balasubramanian:2024ghv, Nandy:2024zcd, Balasubramanian:2024lqk, Ambrosini:2024sre, Baiguera:2025dkc} and proposes its generalizations based on inputs from the bulk. The most notable feature of finite $N$ is that the chord number states are no longer orthogonal with each other, as in the case of geodesic length states in JT gravity \cite{Iliesiu:2024cnh, Miyaji:2024ity, Miyaji:2025yvm}. In particular, due to the Cayley-Hamilton theorem, there are only finitely many linearly independent chord number states, implying there are infinitely many \emph{null states} \cite{Marolf:2020xie} and the actions of $q$-oscillators are \emph{no longer linear}. 

In order to construct a new orthonormal chord number basis on which the action of $q$-oscillators makes sense, we will consider the Gram-Schmidt orthogonalization of the chord number states and identify them as the eigenstates of the new chord number operator. As the chord number states are expressed via $q$-deformed Hermite polynomials, we will show the \emph{exact equivalence} between this newly constructed chord number operator with the ``reference state'' $|0\rangle=|\text{TFD}(0)\rangle$ and the Krylov state complexity operator for the initial state $|\text{TFD}(0)\rangle$, in section \ref{section:finiteN}. Since the Krylov complexity displays the ``ramp-slope-plateau'' behavior \cite{Balasubramanian:2022tpr, Balasubramanian:2022dnj, Nandy:2024zcd, Balasubramanian:2024lqk}, by using this equivalence we can show that the chord number expectation value displays the same behavior, as the geodesic length in JT gravity displays the plateau at late times \cite{Miyaji:2024ity}. We will also demonstrate the ``ramp-slope-plateau'' behavior in explicit examples in section \ref{section:Gaussian}. From this result, we can reach the conclusion that at late times the length of ER bridge saturates, interpreted as the equal superposition of black holes and white holes that have expanding and contracting interior \cite{Susskind:2015toa, Stanford:2022fdt, Miyaji:2024ity, Balasubramanian:2024lqk}. This phenomenon is suggested to be related to firewall \cite{Almheiri:2012rt,Almheiri:2013hfa,Marolf:2013dba,Susskind:2015toa,Mathur:2009hf}, see \cite{Stanford:2022fdt, Blommaert:2024ftn}.

More generally, we will construct a family of finite $N$ chord number operators, by considering the ``EFT for gravitational observables'' version of the ``non-isometric code'' \cite{Akers:2021fut, Akers:2022qdl, Antonini:2024yif}. Namely, we will construct a class of chord number operators for arbitrary reference chord number state $|R\rangle$, such that their eigenstates approximately realize a representation of the $q$-deformed oscillator algebra near the reference state. As the non-isometric code maintains the overlaps of ``all states of sub-exponential complexity'' at the leading order in $1/L$ expansion, the overlaps of chord number states around the reference states are maintained at the leading order. In this more general case, the chord number operator ceases to be identical to the original Krylov state complexity operator, leading to an interesting generalizations. The notable feature of these new chord number operators is that they can reproduce the semi-classical dynamics around the reference state $|R\rangle$, as the same phenomenon was considered for JT gravity in \cite{Miyaji:2024ity}. We will perform explicit, detailed analysis in section \ref{section:Gaussian}.

The above reference state dependence raises an interesting puzzle that the qualitative feature is drastically dependent on the reference state, where similar phenomenon was also observed in \cite{Balasubramanian:2022gmo}, intimately related to the black hole information paradox and null states \cite{Marolf:2020xie}. We discuss this problem further in section \ref{section:finiteN}. We will also discuss this issue in relation to the Poincare recurrence in the section \ref{section:discussion} with comments on future problems.


\section{ETH Matrix Model for Double-scaled SYK}\label{section:ETHmatrixmodel}


In this section, we will explain the strict large $N$ DSSYK model and its ETH matrix model for finite $N$. In particular, we will explain the construction of chord number states, the baby universe in DSSYK, and the discrete Weil-Petersson volume associated with the matrix model.

\subsection{Large $N$ DSSYK model}

Let us review the structure of the DSSYK model at large $N$, which we would like to reproduce from the ETH matrix model at strict large $N$. We denote the Hilbert space dimension of the DSSYK model as
\begin{equation}
    L:=2^{N/2}.
\end{equation}
We will consider the tensor product $\mathcal{H}_{L}\otimes\mathcal{H}_{R}$ of two DSSYK Hilbert spaces, and restrict our attention to the diagonal part $\mathcal{H}:=\text{span}\lbrace |E_i^{L}\rangle|E_i^{R}\rangle\in\mathcal{H}_{L}\otimes\mathcal{H}_{R}\rbrace$, whose dimension is again $L$. The two-sided Hamiltonian $H=H_L+H_R$ of DSSYK at large $N$ is given by the $q$-deformed oscillators $A_\pm$ with $A_-=A_+^{\dagger}$,
\begin{equation}
\begin{aligned}
 H=\frac{E_0}{2}(A_++A_{-}),\quad E_0=\frac{2}{\rt{1-q}}.
\end{aligned} 
\end{equation}
The $q$-deformed oscillator $A_\pm$ obeys
\begin{equation}
\begin{aligned}
 A_-A_+-qA_+A_-&=1-q,\\
 A_+A_-&=1-q^{\hat{N}}.
\end{aligned} 
\end{equation}
The chord number operator $\hat{N}$ is Hermitian and the eigenstates $\hat{N}|n\rangle=n|n\rangle$ form an orthonormal basis. The action of $A_\pm$ on the chord number states is
\begin{equation}
\begin{aligned}
 A_+|n\ket=\rt{1-q^{n+1}}|n+1\ket,\quad
A_-|n\ket=\rt{1-q^n}|n-1\ket,
\end{aligned} 
\end{equation}
thus the chord number state can be expressed as
\begin{equation}
\begin{aligned}
 |n\ket=\frac{A_+^n}{\rt{(q;q)_n}}|0\ket,
\end{aligned} 
\end{equation}
where we introduced $q$-Pochhammer symbol $ (a;q)_n:=\Pi_{k=0}^{n-1}(1-aq^k)$. The commutation relation with the chord number operator is
\begin{equation}
    [A_+,~\hat{N}]=-A_+,~[A_-,~\hat{N}]=A_-.
\end{equation}
The chord number basis is identical to the Krylov basis of DSSYK at strict large $N$ with an initial state $|0\rangle$ \cite{Lin:2022rbf}. The Krylov basis with this initial state is defined by the Gram-Schmidt orthogonalization of the ordered basis $H^n|0\rangle (n=0,1,\cdots)$. Since $H^n|0\rangle$ can be spanned by $|m\rangle~(0\leq m\leq n)$, it is clear that the Krylov basis and the chord number basis coincide in this case.

The bulk geometry emerges out of the chord diagram constructed from the above algebraic structure, without explicitly referring to the structure of the original DSSYK model. The correspondence is
\begin{equation}\label{eq:chorddiagram}
    \imineq{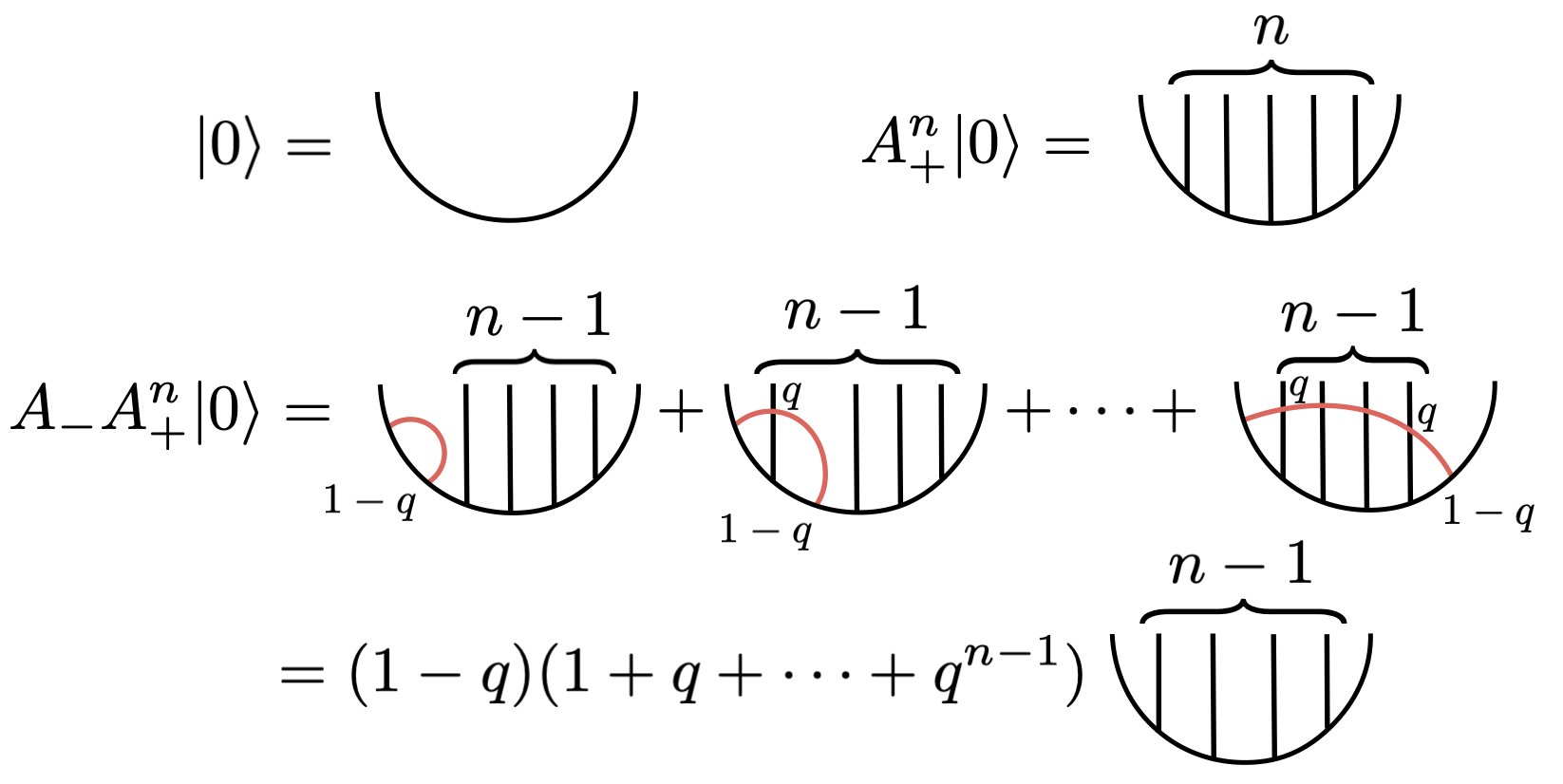}{25}
\end{equation}
Here, each pair of chords can intersect only once, and we multiply by $q$ for each intersection and $(1-q)$ for each contraction. The inner product can be described by the sum over chord diagrams, whose chords are constrained to be paired between bra and ket
\begin{equation}\label{eq:large-L-inner}
    \imineq{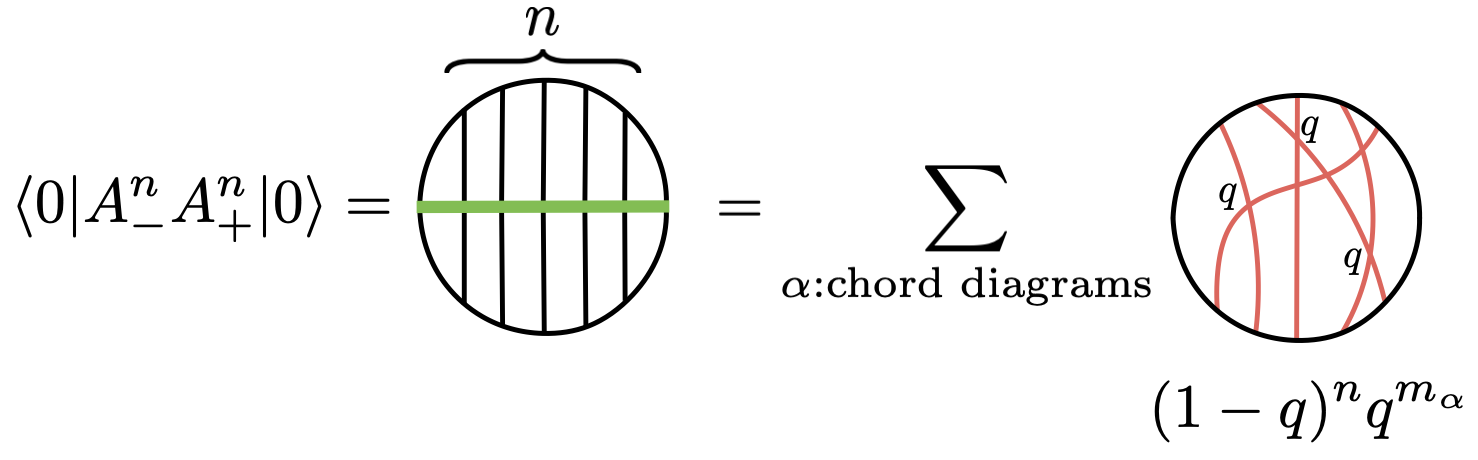}{17},
\end{equation}
here $m_\alpha$ is the number of the intersections in the given chord diagram $\alpha$. Note in particular the diagram for $\langle 0|A_-^nA_+^m|0\rangle$ is zero for $n\neq m$ since such pairing is impossible. 

The disk amplitude of the DSSYK model is
\begin{equation}
    \Tr_{\mathcal{H}}[e^{-\beta H}]=\langle 0|e^{-\beta H}|0\rangle=\int_0^\pi~\frac{d\theta}{2\pi} \mu(\theta)e^{-\beta E(\theta)},
\end{equation}
where
\begin{equation}
    \mu(\th)=(q,e^{\pm2\ri\th};q)_\infty:=(q;q)_\infty(e^{2\ri\th};q)_\infty(e^{-2\ri\th};q)_\infty,
\end{equation}
is the density of states in the $\theta$ coordinate. Here we label the energy spectrum of the model by
\begin{equation}
    E(\theta)=E_0\cos\theta,~(0\leq \theta\leq\pi).
    \label{eq:E-th}
\end{equation}
We normalize the energy eigenstate $|\theta\rangle$ by
\begin{equation}
\begin{aligned}
\bra\th|\th'\ket=\frac{2\pi}{\mu(\th)}\cob(\th-\th'),
\end{aligned}
\label{eq:norm-nth} 
\end{equation}
then the overlap with the chord number state is given by the $q$-Hermite polynomial
\begin{equation}
\begin{aligned}
 \bra n|\th\ket=\frac{H_n(\cos\th|q)}{\rt{(q;q)_n}},
\end{aligned} 
\end{equation}
where
\begin{equation}
\begin{aligned}
 H_n(\cos\th|q)=\sum_{k=0}^n \qb{n\\k}e^{(n-2k)\ri\th},~\qb{n\\k}=\frac{(q;q)_n}{(q;q)_k(q;q)_{n-k}}.
 \end{aligned}
\label{eq:Hn} 
\end{equation}
Thus, we have the resolution of identity
\begin{equation}
\begin{aligned}
 \id=\sum_{n=0}^\infty |n\ket\bra n|=\int_0^\pi\frac{d\th}{2\pi}\mu(\th)
|\th\ket\bra\th|.
\end{aligned} 
\end{equation}
We consider the dynamics of the thermofield double state
\begin{equation}
    |\text{TFD}(\beta,t)\rangle:=\int_0^\pi \frac{d\theta}{2\pi} \mu(\theta)e^{-(\beta+it)E(\theta)}|\theta\rangle,
\end{equation}
here we note that we use a convention for the inverse temperature $\beta$ instead of $\beta/2$. Since $\bra 0|\th\ket=1$ we have
\begin{equation}
    |\text{TFD}(\beta=0,t=0)\rangle=|0\rangle.
\end{equation}
Let us evaluate the probability distribution of the chord number of the TFD state
\begin{equation}
\begin{split}
    P_n(\beta,t)
    &:=\frac{1}{\Tr_{\mathcal{H}}[e^{-\beta H}]}\langle\text{TFD}(\beta,t)|n\rangle\langle n|\text{TFD}(\beta,t)\rangle\\
    &=\frac{1}{\Tr_{\mathcal{H}}[e^{-\beta H}]}\int_0^\pi d\theta d\theta'\mu(\theta)\mu(\theta')e^{-(\beta-it)E(\theta)-(\beta+it)E(\theta')}\frac{H_n(\cos\th|q)}{\rt{(q;q)_n}}\frac{H_n(\cos\th'|q)}{\rt{(q;q)_n}}.
\end{split}
\end{equation} 
This probability distribution at the strict large $N$, for the JT gravity was considered in \cite{Stanford:2022fdt, Blommaert:2024ftn, Iliesiu:2024cnh, Miyaji:2024ity, Miyaji:2025yvm} and the DSSYK in \cite{Xu:2024hoc, Xu:2024gfm}. One of the primary goals of this paper is to study the finite $N$ version of this distribution.

The velocity operator $\pi_N$ for the chord number operator $\hat{N}$ is defined by
\begin{equation}
\begin{aligned}
 \pi_N=\ri[H,N]=\frac{\ri E_0}{2}(A_{-}-A_{+}).
\end{aligned} 
\end{equation}
This operator describes the speed of expansion of the chord number. The velocity operator $\pi_N$ acts on wavefunction $H_n$ as
\begin{equation}
\begin{aligned}
 \pi_N H_n&=\frac{\ri E_0}{2}\Bigl[(1-q^n)H_{n-1}-H_{n+1}\Bigr].
\end{aligned} 
\end{equation}
In the $n$ regime where $q^n\ll1$, this is approximated as
\begin{equation}
\begin{aligned}
 \pi_N H_n&\approx \frac{\ri E_0}{2}\Bigl[H_{n-1}-H_{n+1}\Bigr].
\end{aligned} 
\end{equation}
In this limit, the wavefunction can be approximated by the sum of the plane-wave solutions $H_n^\pm\sim e^{\pm\ri n\th}$ \footnote{The full expression is
\begin{equation}
    H_n^\pm=\frac{e^{\pm in\theta}}{(e^{\mp2i\theta};q)_\infty}\sum_{k=0}^{\infty}
    \frac{q^{k(n+1)}}{(q;q)_k(e^{\pm2i\theta}q;q)_k},~H_n=H_n^++H_n^-.
\end{equation}

}, and the eigenvalue of $\pi_N$ is given by
\begin{equation}
\begin{aligned}
 \pi_Ne^{\pm\ri n\th}=\pm E_0\sin\th e^{\pm\ri n\th}.
\end{aligned} 
\label{eq:plane}
\end{equation}
This is consistent with the stationary phase approximation of the integral
$\int d\th e^{-\ri n\th-\ri tE(\th)}f(\th)$
in the limit $n,t\gg1$
\begin{equation}
\begin{aligned}
 \frac{\del}{\del\th}\bigl[n\th+tE(\th)\bigr]=0\quad\Rightarrow\quad
n=-tE'(\th)=tE_0\sin\th,
\end{aligned} 
\label{eq:stationary}
\end{equation}
where we used $E(\th)=E_0\cos\th$ \footnote{The same approximation
can be used for the JT gravity case $\int dk e^{\ri k\ell-\ri tE(k)}f(k)$ with $E(k)=k^2$
\begin{equation}
\begin{aligned}
 \frac{\del}{\del k}\bigl[k\ell-tE(k)\bigr]=0\quad\Rightarrow\quad
\ell=tE'(k)=2t k.
\end{aligned} 
\end{equation}
where $\th$ and $k$ are related by $\th=\pi-\la k$ \cite{Berkooz:2018jqr}.}.


\subsection{ETH matrix model for DSSYK}
We consider the $L\times L$ hermitian one-matrix model, dubbed ETH matrix model, which reproduces the large $N$ DSSYK at the leading order in large $L=2^{N/2}$ expansion \cite{Jafferis:2022uhu, Jafferis:2022wez, Okuyama:2023aup, Okuyama:2023kdo}
\begin{equation}
    \langle\prod_{i=1}^nZ(\beta_i)\rangle=\frac{1}{\mathcal{Z}}\int~dH~e^{-L\Tr[V(H)]}\prod_{i=1}^n\Tr[e^{-\beta_iH}],
    \label{eq:ETH}
\end{equation}
where $\cZ=\int dH~e^{-L\Tr[V(H)]}$.
If we include the effect of matter operators,  the ETH matrix model is given by a two-matrix model \cite{Jafferis:2022uhu, Jafferis:2022wez},
but we have ignored the matter operators for simplicity. Then the ETH matrix model in \cite{Jafferis:2022uhu, Jafferis:2022wez} reduces to the one-matrix model
in \eqref{eq:ETH}.
The potential $V(H)$ is determined by matching the density of states $\mu(\th)$ of DSSYK
and the genus-zero eigenvalue density $\rho_0(E)$ of the matrix model \eqref{eq:ETH}
\begin{equation}
    \frac{d\th}{2\pi}\mu(\th)=dE\rho_0(E),
    \label{eq:match}
\end{equation}
where $E$ and $\th$ are related by \eqref{eq:E-th} at the disk level.
The explicit form of $V(H)$ is given by
\begin{equation}
V(H)=\sum_{n=1}^\infty\frac{(-1)^{n-1}}{n}q^{\hf n^2}(q^{\hf n}+q^{-\hf n})T_{2n}(H/E_0),
\end{equation}
where $T_n(x)$ denotes the Chebyshev polynomial of the first kind.

We note that this model \emph{differs} from the large $N$ expansion of the SYK model, as the multi-boundary connected correlation function of the SYK model
decays as series of $1/\log L$ instead of $1/L$ expansion \cite{Berkooz:2020fvm}. 
We consider the large $L$ expansion of the ETH matrix model \eqref{eq:ETH} in its own right.

As a single-cut one-matrix model, the connected part of the resolvent correlation function can be expanded as
\begin{equation}
    \langle\prod_{i=1}^n\frac{1}{x_i+H}\rangle_{\text{connected}}=\sum_{g=0}^{\infty}L^{2-2g-n}W_{g,n}(x_1,\cdots,x_n).
\end{equation}
We introduce
\begin{equation}
    \omega_{g,n}(z_1,\cdots,z_n)=W_{g,n}(x(z_1),\cdots,x(z_n))dx(z_1)\cdots dx(z_n),
\end{equation}
where
\begin{equation}
    dx(z)=\frac{E_0}{2}(1-\frac{1}{z^2})dz.
\end{equation}
Then we have the Eynard-Orantin topological recursion relation
\begin{equation}
\begin{split}
    \omega_{g,n+1}(z_0,J)
    &=\sum_{\alpha=\pm 1}\underset{z=\alpha}{\text{Res}}\Big[K(z_0,z)
    \\&\times
    \Big[\omega_{g-1,n+2}(z,\bar{z},J)+\underset{(h,I)\neq (0,\emptyset),(g,J)}{\sum_{h=0}^g\sum_{I\subset J}}\omega_{h,|I|+1}(z,I)\omega_{g-h,1+n-|I|}(\bar{z},J-I)\Big]\Big],
\end{split}
\end{equation}
with the initial condition
\begin{equation}
    \omega_{0,2}(z_1,z_2)=\frac{dz_1dz_2}{(z_1-z_2)^2},
\end{equation}
and the recursion kernel
\begin{equation}
    K(z_0,z)=-\frac{\int_{z'=\bar{z}}^z\omega_{0,2}(z_0,z')}{4y(z)dx(z)}.
\end{equation}
We can express
\begin{equation}
    \omega_{g,n}(z_1,\cdots,z_n)=\sum_{b_1,\cdots,b_n\in\mathbb{Z}_+}N_{g,n}(b_1,\cdots,b_n)\prod_{i=1}^nb_iz_i^{b_i-1}dz_i.
\end{equation}
Then we can arrive at
\begin{equation}
    Z_{g,n}(\beta_1,\cdots,\beta_n)=\sum_{b_1,\cdots,b_n\in\mathbb{Z}_+}N_{g,n}(b_1,\cdots,b_n)\prod_{i=1}^nb_iZ_{\text{Trumpet}}(\beta_i,b_i),
    \label{eq:WP-trumpet}
\end{equation}
where $Z_{\text{Trumpet}}(\beta,b)$ is the trumpet partition function of DSSYK \cite{Jafferis:2022wez,Okuyama:2023byh}
\begin{equation}
    Z_{\text{Trumpet}}(\beta,b)=\int_0^\pi\frac{d\th}{2\pi}e^{-\bt E(\th)}2\cos(b\th)=I_b(-\beta E_0),
\end{equation}
and $N_{g,n}(b_1,\cdots,b_n)$ in \eqref{eq:WP-trumpet} is the discrete version of the Weil-Petersson volume 
introduced by Norbury and Scott \cite{norbury2013polynomials}. We note that this a discretization of WP volume and there could be other possible discretizations.

\subsection{Trumpet and baby universe operator}
One direct geometric approach to understand the saturation of the geodesic length at late time is to consider baby universe emissions. Although we will not use this direct approach to understand the late time saturation in this paper, we will briefly explain its basic properties in this subsection.

We can rewrite the trumpet partition function as
\begin{equation}
\begin{aligned}
 Z_{\text{trumpet}}(\bt,b)=\bra 0|e^{-\bt T}\cB_b|0\ket
\end{aligned} 
\end{equation}
using the baby universe operator $\cB_b$ \cite{Penington:2023dql}
\begin{equation}
\begin{aligned}
 \cB_b=\int_0^\pi\frac{d\th}{2\pi}2\cos(b\th)|\th\ket\bra\th|,~b\in\mathbb{Z}_{\geq0}.
\end{aligned} 
\end{equation}
Let us consider the matrix element
$\bra n|\cB_b|m\ket$ of baby universe operator.
Using \eqref{eq:Hn} we find
\begin{equation}
\begin{aligned}
 \bra n|\cB_b|m\ket&=\int_0^\pi\frac{d\th}{2\pi}2\cos(b\th)
\frac{H_n(\cos\th|q)}{\rt{(q;q)_n}}
\frac{H_m(\cos\th|q)}{\rt{(q;q)_m}}\\
&=\frac{1}{\rt{(q;q)_n(q;q)_m}}\int_0^{2\pi}\frac{d\th}{2\pi}e^{-b\ri\th}
\sum_{k=0}^n\qb{n\\k}e^{(n-2k)\ri\th}
\sum_{j=0}^m\qb{m\\j}e^{(m-2j)\ri\th}\\
&=\frac{1}{\rt{(q;q)_n(q;q)_m}}
\sum_{\substack{k+j=\frac{n+m-b}{2}\\0\leq k\leq n,\,0\leq j\leq m}}
\qb{n\\k}\qb{m\\j}.
\end{aligned} 
\end{equation}
Here the $q$-binomial should be understood as
\begin{equation}
\begin{aligned}
 \qb{n\\k}=0,\quad(k<0,~~k>n).
\end{aligned} 
\end{equation}
Alternatively, using the linearization formula of $q$-Hermite polynomials
\begin{equation}
\begin{aligned}
 H_n(x|q)H_m(x|q)=\sum_{k=0}^{\min(n,m)}\frac{(q;q)_n(q;q)_m}{(q;q)_{n-k}(q;q)_{m-k}(q;q)_k}H_{n+m-2k}(x|q),
\end{aligned} 
\end{equation} 
the baby universe amplitude $\bra n|\cB_{a}|m\ket$ is expanded as
\begin{equation}
\begin{aligned}
 \bra n|\cB_{a}|m\ket=\sum_{k=0}^{\min(n,m)}
\bra 0|\cB_{a}|n+m-2k\ket
\rt{\qb{n\\k}\cdot\qb{m\\k}\cdot\qb{n+m-2k\\n-k}}.
\end{aligned} 
\label{eq:nbm-expand}
\end{equation}
Note that 
\begin{equation}
\bra 0|\cB_{a}|2k\ket=\left\{
\begin{aligned}
 &\frac{\rt{(q;q)_{2k}}}{(q;q)_{k-a/2}(q;q)_{k+a/2}},\quad &&
(0\leq a\leq 2k,~a=\text{even}),\\
&0,\quad &&(\text{otherwise}).
\end{aligned} \right.
\label{eq:0b2k}
\end{equation}
\begin{figure}[t]
\centering
\subcaptionbox{$q=0.1$ \label{sfig:mat}}{\includegraphics[width=0.4\linewidth]{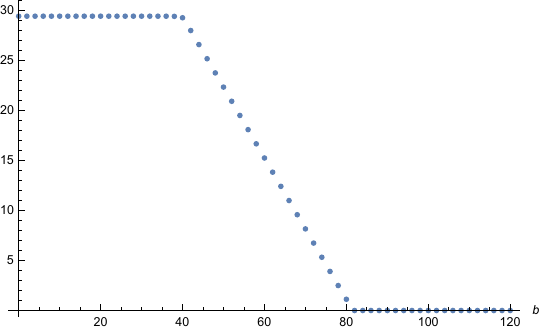}}
\hskip6mm
\subcaptionbox{$q=10^{-4}$ \label{sfig:mat2}}{\includegraphics[width=0.4\linewidth]{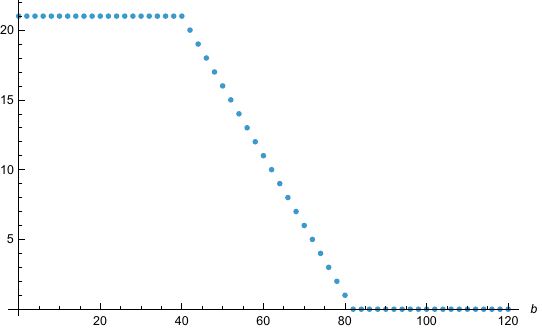}}
  \caption{
Plot 
of $\bra n|\cB_b|m\ket$ as a function of $b$.
In this figure, we have set $n=60,m=20$, and \subref{sfig:mat} $q=0.1$
and \subref{sfig:mat2} $q=10^{-4}$. 
}
  \label{fig:nbm}
\end{figure}
When $q=0$, the matrix element is particularly simple and is piecewise linear in $b$. For $n\geq m$, we have 
\begin{equation}
\bra n|\cB_{a}|m\ket=\left\{
\begin{aligned}
 &m+1,
 &&
(0\leq b\leq n-m),\\
&\frac{n+m+2-b}{2},&&(n-m\leq b\leq n+m+1),
\\& 0,&&(n+m+2\leq b).
\end{aligned} \right.
\label{eq:0b2k}
\end{equation}
This formula indicates the emission of the baby universe with $b\leq |n-m|$ is most probable. As we can see from Figure~\ref{fig:nbm},
there are two special values of $b$
for the matrix element $\bra n|\cB_b|m\ket$ with a fixed $(n,m)$
\begin{equation}
\begin{aligned}
 b=|n-m|,~n+m.
\end{aligned} 
\end{equation}
For $b<|n-m|$, the matrix element is approximately constant
\begin{equation}
\begin{aligned}
 \bra n|\cB_b|m\ket\approx \bra n|\cB_0|m\ket,\qquad(0\leq b\leq|n-m|),
\end{aligned} 
\end{equation}
and it vanishes for $b>n+m$
\begin{equation}
\begin{aligned}
 \bra n|\cB_b|m\ket=0,\qquad(b>n+m).
\end{aligned} 
\label{eq:mat-vanish}
\end{equation}
The special value $b=|n-m|$ corresponds to the ``wormhole shortening''
discussed in \cite{Saad:2019pqd,Stanford:2022fdt}.


\section{ETH matrix model and the length of BH interior at finite $N$}\label{section:finiteN}

In this section, we first construct the chord number states $|n\rangle$ at finite $N$. We show that these states are not orthogonal due to finite $N$ effects. In particular, the chord number states $|m\rangle$ with $m\geq L$ are superpositions of $L$ smaller chord number states $|n\rangle (0\leq n\leq L-1)$ due to the Cayley-Hamilton theorem. By performing orthogonalizations to various chord numbers to obtain an over-complete ordered basis, we construct a family of new chord number basis that probes the bulk differently. In particular, we show that the particular basis $|n\rangle (0\leq n\leq L-1)$ leads to the Krylov basis with initial state $|0\rangle$ non-perturbatively for all $q$. 


\subsection{Non-orthogonal finite $L$ chord number state without averaging}

Let us consider the ETH matrix model for DSSYK. We denote the wavefunction of the chord number state as
\begin{equation}
    W_n(E):=\bra n|\th\ket=\frac{H_n(E/E_0|q)}{\rt{(q;q)_n}}.
    \label{eq:Wn}
\end{equation}
By construction, $W_n(E)$ is orthonormal with respect to $\rho_0(E)$
in \eqref{eq:match}  
\begin{equation}
\begin{aligned}
 \bra n|m\ket=\int_{-\infty}^\infty dE\rho_0(E)W_n(E)W_m(E)=\cob_{n,m}.
\end{aligned} 
\end{equation}
This wavefunction $W_n(E)$ is originally introduced as the orthonormal polynomial at the disk level.
We continue to use
$W_n(E)$ in the ETH matrix  at finite $L$ and extend the support of $W_n(E)$ to the whole real axis $E\in\mathbb{R}$.
This extension can be defined unambiguously since $W_n(E)$ is a polynomial in $E$.

So far we have discussed a large $L$ chord number state. At finite $L$, the bulk geometry can have trumpets and higher genus geometries attached to these trumpets. In particular, the chord number states at finite $L$ are no longer orthogonal to each other, similar to the geodesic length states in JT gravity \cite{Iliesiu:2024cnh, Miyaji:2024ity, Miyaji:2025yvm}. In the following, we will construct the chord number states and the bulk geometry based on the DSSYK algebra and associated chord diagrams. The main differences from the strict large $L$ are that the $q$-deformed oscillators $A_+,~A_-,~\hat{N}$ are no longer linear operators but rather maps acting on the Hilbert space, and the structure of the inner product in chord diagram receives finite $L$ correction. Yet the $q$-deformed oscillators $A_+$, $A_-$ do satisfy the relation
\begin{equation}\label{eq:commutation}
\begin{aligned}
 &H=\frac{E_0}{2}(A_++A_{-}),\quad E_0=\frac{2}{\rt{1-q}},\\
 & A_-A_+-qA_+A_-=1-q,\quad A_+A_-=1-q^{\hat{N}}\\
 &[A_+,~\hat{N}]=-A_+,~[A_-,~\hat{N}]=A_-.
 \end{aligned} 
\end{equation}
when they act on the chord number states. The reason why they are not linear operators is that they do not commute with the Cayley-Hamilton equality, as we will explain in the next subsection. 

Let us explain the details of the construction of the chord number states and the corresponding bulk at finite $L$ in the following. For finite $L$, we define the chord number state for a single instance of the ensemble as
\begin{equation}\label{eq:chordnumberstates}
\begin{split}
    |n\rangle&=\sum_{i=1}^LW_n(E_i)|E_i\rangle
    =W_n(H)|0\rangle,~(n\in\mathbb{Z}_{\geq0}),
\end{split}
\end{equation}
where $|E_i\rangle$ is the eigenstate of the random matrix $H$ in the ETH matrix model \eqref{eq:ETH}
\begin{equation}
    H|E_i\ket=E_i|E_i\ket,\quad \bra E_i|E_j\ket=\cob_{ij},
\end{equation}
and the wavefunction is written as $W_n(E)=\bra E|n\ket$.
Note that since $W_0(E)=1$, the chord number $0$ state is again the thermofield double state at infinite temperature
\begin{equation}
    |0\rangle=\sum_{i=1}^L|E_i\rangle.
    \label{eq:0-sum}
\end{equation}
Here the thermofield double state is defined by
\begin{equation}
    |\text{TFD}(\beta)\rangle=\sum_{i=1}^Le^{-\beta E_i/2}|E_i\rangle,
\end{equation}
which is not normalized.
On these chord number states, we define the action of the $q$-deformed oscillators as 
\begin{equation}\label{eq:Aplusminus}
\begin{aligned}
 A_+|n\ket=\rt{1-q^{n+1}}|n+1\ket,\quad
A_-|n\ket=\rt{1-q^n}|n-1\ket,
\end{aligned} 
\end{equation}
which leads to
\begin{equation}
\begin{aligned}
 &|n\ket=\frac{A_+^n}{\rt{(q;q)_n}}|0\ket,~\hat{N}|n\rangle=n|n\rangle.
 \end{aligned} 
\end{equation}
As advertised, the commutation relations \eqref{eq:commutation} do hold on these chord number states. Furthermore, the bulk geometry in terms of chord diagrams can be obtained as 
\begin{equation}
    \imineq{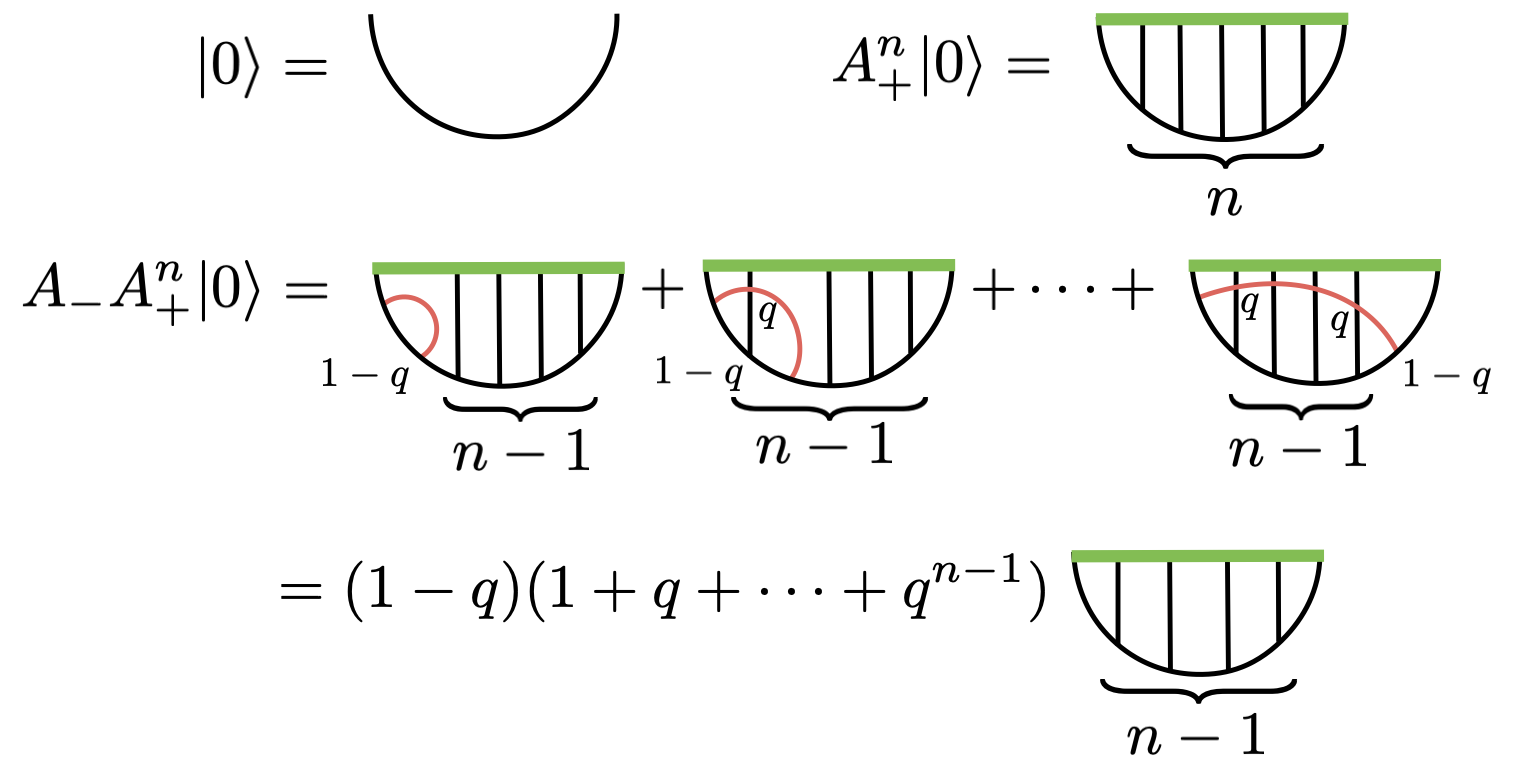}{28}.
\end{equation}
The primary difference from the \eqref{eq:chorddiagram} is the structure of bulk legs schematically depicted as a green timeslice, whose structure depends on the microscopic information of the Hamiltonian. When we take the inner product of the chord number state, unlike the strict large $L$ case, it is unclear whether we can write such inner product in terms of chord diagrams even when we take ensemble averaging. One speculation would be that the timeslice contain a superposition of $n$ chords with $0\leq n\leq L-1$ with microscopic coefficients. We leave these questions to future investigation. While complete chord diagram interpretation is unclear, we know the explicit expression for the inner product, and the finite $L$ effects render the orthogonal basis into a non-orthogonal set of states. Indeed, the finite $L$ effects modifies the inner product $\bra n|m\ket$ to
\begin{equation}
\begin{aligned}
\mathbb{E}[\bra n|m\ket]=\frac{1}{L}\int_{-\infty}^\infty dE~\rho(E)W_n(E)W_m(E),
\end{aligned} 
\label{eq:modified-inner}
\end{equation}
which is no longer orthogonal. Here the eigenvalue density $\rho(E)$ is given by the diagonal part
of the Christoffel-Darboux
(CD) kernel
\begin{equation}
\begin{aligned}
 \rho(E)=K(E,E).
\end{aligned} 
\end{equation}
The CD kernel $K(E_1,E_2)$ is defined by
\begin{equation}
\begin{aligned}
 K(E_1,E_2)&=e^{-\frac{L}{2}V(E_1)-\frac{L}{2}V(E_2)}\sum_{n=0}^{L-1}\frac{P_n(E_1)P_n(E_2)}{h_n}\\
&=\frac{e^{-\frac{L}{2}V(E_1)-\frac{L}{2} V(E_2)}}{h_{L-1}}
\frac{P_L(E_1)P_{L-1}(E_2)-P_{L-1}(E_1)P_{L}(E_2)}{E_1-E_2}
\end{aligned} 
\end{equation}
where $P_n(E)=E^n+\cdots$ denotes the $n$-th orthogonal polynomial
\begin{equation}
\begin{aligned}
 \int dE e^{-LV(E)}P_n(E)P_{m}(E)=h_n\cob_{n,m},~(n,m\in\mathbb{Z}_{\geq 0}).
\end{aligned} 
\label{eq:def-P}
\end{equation}
In general, this polynomial is distinct from $W_n(E)$. Using the above expression, we have
\begin{equation}
\begin{aligned}
 \rho(E)=K(E,E)=\frac{e^{-LV(E)}}{h_{L-1}}
\Big(P_L'(E)P_{L-1}(E)-P_{L-1}'(E)P_{L}(E)\Big).
\end{aligned} 
\end{equation}
We note that in the large $L$ limit, $\rho(E)$ is expanded as
\begin{equation}
\begin{aligned}
 \rho(E)=\sum_{g=0}^\infty L^{1-2g}\rho_g(E).
\end{aligned} 
\end{equation}
Thus $\rho(E)$ is approximated by $L\rho_0(E)$ for large $L$.

\subsection{Null States from Cayley-Hamilton Theorem}

The inner product at fixed $H$ is written as
\begin{equation}
\begin{aligned}
 G_{nm}(H)=\langle n|m\rangle=\sum_{i=1}^LW_n(E_i)W_m(E_i)=\Tr\Bigl[W_n(H)W_m(H)\Bigr].
\end{aligned} 
\label{eq:G-def}
\end{equation}
$H$ is an $L\times L$ matrix in the ETH matrix model \eqref{eq:ETH}, while the matrix $G_{nm}(H)$ is a $\infty\times\infty$ matrix. It turns out that the chord number state is highly redundant and most of the states are linear combinations of other states. Indeed, the rank of the matrix $G_{nm}(H)$ is equal to $L$. This implies that only $L$ chord number states are linearly independent.

To see this, we use the Cayley-Hamilton theorem, which states that an arbitrary $L\times L$ matrix $H$
obeys the identity
\begin{equation}
\begin{aligned}
 H^L+\sum_{i=1}^Lc_i H^{L-i}=0,
\end{aligned} 
\label{eq:CH-thm}
\end{equation}
where
\begin{equation}
\begin{aligned}
 c_i=\sum_{\sum_\ell\ell k_\ell=i}\prod_{\ell=1}^i\frac{1}{k_\ell!}\left[-\frac{1}{\ell}
\Tr H^\ell\right]^{k_\ell}.
\end{aligned} 
\end{equation}
For instance,
\begin{equation}
\begin{aligned}
 c_1=-\Tr H,\quad c_2=\hf(\Tr H)^2-\hf\Tr H^2.
\end{aligned} 
\end{equation}
Thus we can always rewrite a polynomial of $H$ of degree greater than or equal to $L$ in terms of a polynomial of degree less than $L$. This implies that for any $n\geq L$, there exist $d_m(n)$ such that
\begin{equation}
    |n\rangle=\sum_{m=0}^{L-1}d_m(n)|m\rangle.
\end{equation}
In other words, the set of chord number states $|n\rangle~(0\leq n\leq L-1)$ can express arbitrary chord number state.

From \eqref{eq:CH-thm}, one can easily show that the determinant of the Gram matrix $G_{nm}(H)_{0\leq n,m\leq K-1}$ vanishes
when  the size $K$ of $G_{nm}$ is larger than the dimension $L$ of the Hilbert space
\begin{equation}
\det G_{nm}(H)_{K\times K}=0,\quad (K>L).
\end{equation}
The vanishing of the determinant of the Gram matrix implies the existence of null states.
Our argument for the existence of null states in the ETH matrix model of DSSYK is much simpler than the JT gravity case discussed in
\cite{Iliesiu:2024cnh}.
Our argument is based on the following two facts: (i) the wavefunction $W_n(H)$ is an $n$-th order polynomial in $H$, (ii)
$H^L$ is written as a linear combination of $H^l~(l=0,\cdots,L-1) $.
The first point (i) is a peculiar feature of DSSYK which does not have a counterpart in JT gravity:
The fixed length wavefunction in JT gravity, given by the Bessel function, is not a polynomial in energy.
The second point (ii) is a property of the finite size matrix, which is responsible for the existence of null states
and makes the Hilbert space finite-dimensional.
This is along with the similar line as the \emph{holographic covering} in the discussion of fortuitous BPS states in $\cN=4$ super Yang-Mills, where one initiates with $N=\infty$ Hilbert space and observable algebra and modding them out by trace relations of finite size matrices \cite{Chang:2024zqi}.

We can also show that the determinant of the Gram matrix of size $L$ is non-zero
\begin{equation}
\begin{aligned}
 \det (G_{nm}(H))_{0\leq n,m\leq L-1}=\cN\prod_{i<j}(E_i-E_j)^2,
\end{aligned} 
\label{eq:detMN}
\end{equation}
where $\cN$ is a $q$-dependent normalization constant
\begin{equation}
\cN=\prod_{n=0}^{L-1}\frac{(1-q)^n}{(q;q)_n}.
\end{equation}
From this expression, we can conclude that as long as the energy spectrum is non-degenerate, the set of chord number state $|n\rangle~(0\leq n\leq L-1)$ is linearly independent, and thus forms the basis of the finite $L$ Hilbert space. To show this, we observe that
\begin{equation}
\begin{aligned}
 G_{nm}(H)&=\Tr \bigl[W_n(H)W_m(H)\bigr]\\
&=\sum_{k=0}^{L-1}W_n(E_{k})W_m(E_{k})\\
&=(VV^T)_{nm}
\end{aligned} 
\label{eq:VV}
\end{equation}
where
\begin{equation}
\begin{aligned}
 V_{nk}=W_n(E_{k}),\quad (0\leq n\leq L-1,~1\leq k\leq L).
\end{aligned} 
\end{equation}
Using $\det V=\det V^T=\cN^{1/2}\prod_{i<j}(E_j-E_i)$, 
one can see that \eqref{eq:detMN} follows from \eqref{eq:VV}
\begin{equation}
\begin{aligned}
 \det (G_{nm}(H))_{0\leq n,m\leq L-1}=\det V\det V^T=\cN\prod_{i<j}(E_i-E_j)^2.
\end{aligned} 
\end{equation}
It is possible to consider a more general set of chord number states to form the basis of the Hilbert space. Namely, we can consider $|n\rangle~(R\leq n\leq R+L-1)$ as the alternative basis. The condition for linear independence in this case is slightly more complicated as we need more conditions on the energy spectrum other than non-degeneracy. In this case, any $n$-th chord number state is a superposition of $|n\rangle~(R\leq n\leq R+L-1)$ with coefficients that are dependent on the microscopic energy spectrum.

\subsubsection*{Non-linearity of $A_{\pm}$ and $\hat{N}$}
A necessary condition for the linearity of an operation is that its action preserves null states. We can show that this condition \emph{does not} hold for $A_{\pm}$ and $\hat{N}$; the action of $A_{\pm}$ does not preserve null states. The simplest example can be given in $q=0$, $L=2$ case. Therefore the action of $A_{\pm}$ and $\hat{N}$ \eqref{eq:Aplusminus} are defined only on the chord number states \eqref{eq:chordnumberstates}, and cannot be consistently extended to their superpositions. In the next subsection, we will explore \emph{linear}, finite $N$ chord number operators whose eigenstates are obtained via orthogonalization of the chord number states. Nonetheless, we will argue that the actions of $q$-oscillators are approximately linear on these newly constructed orthogonal chord number basis, in a restricted sense explained in the following.

\subsection{Finite $L$ chord number operators}

As we have seen, the chord number states we constructed are non-orthogonal, and form an overcomplete basis of the $L$-dimensional non-perturbative Hilbert space. The naive chord number operator
\begin{equation}
    \hat{N}^{\text{naive}}:=\sum_{n\geq 0} n|n\rangle\langle n|,
\end{equation}
is not well-defined and divergent due to the unbounded nature of the chord number. Furthermore, the chord number states $|n\rangle$ are not guaranteed to be eigenstates of $\hat{N}^{\text{naive}}$. The only way to make sense of this operator is to take $e^{-\alpha\hat{N}^{\text{naive}}}~(\alpha>0)$, such that large $n$ contribution is exponentially suppressed. The expectation value of $e^{-\alpha\hat{N}^{\text{naive}}}$ displays the ``dip-ramp-plateau'' behavior as in the spectral form factor and can prove the chaotic spectrum of the Hamiltonian \cite{Miyaji:2025yvm}.

Furthermore, it is not obvious whether it is possible to construct the decomposition of Hilbert space into distinct chord numbers, wormhole velocities, and black hole/white hole, which are at this point classical characterizations. Let us illustrate an example from the minimal string that the classical and quantum descriptions can differ drastically \cite{Maldacena:2004sn}. The minimal string at $p=2$ is dual to one-matrix model, and FZZT brane is described by the determinant operator $\text{det}(x-H)$, where $x$ is the boundary cosmological constant labeling the FZZT brane, which parametrizes the target space. It was found that while the classical target space is a multi-sheeted Riemann surface with a branch cut, the quantum one is just the entire complex plane. Similarly, the density of states has support on a semi-infinite line classically while quantum mechanically it is non-zero everywhere on the real line. 

In our current ETH matrix model, we encounter similar phenomena. Classically, the support of the density of states is an interval $-E_0<E<E_0$. With the inclusion of the $1/L$ effect, the support is no longer the interval but extends to the entire real line. Indeed, the wavefunction $W_n(E)=\bra n|\th\ket$ of $n$-th chord number state is $q$-Hermite polynomial, which is a polynomial of $E$ of degree $n$, that can take non-zero value everywhere except for finite zeros on the real line. Furthermore, since the Hilbert space is finite-dimensional, the $q$-oscillator algebra representation cannot be exact, as it requires infinite dimensional space. Let us consider the canonical conjugate momentum of the chord number operator $[\hat{N},~\hat{P}]=i$ whose eigenstate is $|[p]\rangle:=\sum_{n=0}^\infty e^{in[p]}|n\rangle$ where $[p]$ is the equivalence class $p\sim p+2\pi$. Let us choose the representatives as $-\pi<p<\pi$ and denote $[p]$ as $p$. Since the wavefunction is $\langle n|p\rangle=e^{inp}$, the positive/negative $p$ corresponds to the black hole/white hole states. However, similar to chord number states $|n\rangle$, fixed momentum state $|p\rangle$ are non-orthogonal with each other at finite $L$ and the naive conjugate momentum operator $\hat{P}^{\text{naive}}$ suffers the ill-defined nature and divergence, whose eigenstates are not guaranteed to be $|p\rangle$. The basic reason behind this problem is that the canonical commutation relation cannot be realized in a finite-dimensional Hilbert space.

Another way to phrase this problem is that since the energy spectrum consists of $L$ points, it is impossible to distinguish more than $L$ chord number states. This problem is an alternative way to phrase the Cayley-Hamilton theorem and the non-linearity of the $q$-deformed oscillator in the previous subsection. This leads us to a freedom to \emph{choose} $L$ chord number states $|n\rangle~(n\in \mathcal{S},~|\mathcal{S}|=L)$ in constructing refined chord number operator.

\subsubsection*{EFT for gravitational observables and finite $L$ chord number basis}

We will now describe our proposal to construct finite $L$ chord number operators. Our strategy is to apply orthogonalization to the 
chosen $L$ non-orthogonal chord number basis states $|n\rangle~(n\in \mathcal{S},~|\mathcal{S}|=L)$. Here we emphasize that the number of chord number states is infinite, thus we are now choosing a finite set $\mathcal{S}$. The obtained orthogonal basis $|n^X\rangle~(n^X\in \mathcal{S})$ gives the chord number operator via
\begin{equation}
    \hat{N}^X:=\sum_{n^X\in\mathcal{S}}n^X|n^X\rangle\langle n^X|.
\end{equation}
This operator is manifestly linear, and $|n^X\rangle$ is an eigenstate with eigenvalue $n^X$. 

In this construction, how to choose $\mathcal{S}$ and orthogonalize them is crucial. We look for the chord number operators such that they can be described via EFT for gravitational observables. Thus the criterion we require is:
\begin{center}
    \textit{The constructed chord number states $|n^X\rangle$ realize a representation of the $q$-deformed oscillator algebra 
    approximately near the reference chord number state $|R^X\rangle=|R\rangle$, such that semi-classical gravity can effectively describe the dynamics around the reference state.}
\end{center}
Here we introduced the \emph{reference chord number state} $|R^X\rangle=|R\rangle$, and we require that actions of $A_+,~A_-$ satisfying the commutation relations \eqref{eq:commutation} can be realized approximately as linear operators on constructed chord number states $|n^X\rangle$ when $|n^X-R|$ is sufficiently small.

This criterion is closely related to the ``nonisometric code'' in the entanglement wedge reconstruction for matter EFT at the black hole horizon \cite{Akers:2021fut, Akers:2022qdl, Antonini:2024yif}. Here we applied this formalism to the gravitational observables such as length, as was first considered in \cite{Penington:2023dql, Akers}. In the non-isometric code, it was important to maintain the overlaps of ``all states with sub-exponential complexity'', namely states with a sufficiently small number of excitations on the vacuum. This guarantees that the non-isometric feature cannot be detected without exponentially complex operations. In the current case, constructed chord number states $|n^X\rangle$ with sufficiently small $|n^X-R|$ play the identical role.

In the above construction, one may wonder what is the role played by the orthogonalization. It is indeed crucial to perform orthogonalization. This is because of the non-perturbative non-orthogonal nature of the chord number states. If we did not perform orthogonalization, the action of $\hat{N}^X$ on the neighborhood of the reference state $|R^X\rangle=|R\rangle$ may be perturbed significantly, and the criterion that the approximate action of $q$-deformed oscillator algebra may not be fulfilled.

Let us explain more concretely how we construct the orthogonal basis $|n^X\rangle$ using Gram-Schmidt orthogonalization. We first choose $L$ non-orthogonal chord number states with consecutive chord numbers around the reference state $|R\rangle$. We then assign an ordering, such as
\begin{equation}
    |R\rangle,~|R\pm1\rangle,~|R\pm2\rangle,~\cdots,
\end{equation}
and we apply the Gram-Schmidt (or other procedures of) orthogonalization to this ordered basis. It is clear that the effect of orthogonalization on $|n^X\rangle$ is small when $|n-R|$ is order one, thus we can construct the approximate representation of the $q$-deformed oscillator algebra. We identify such realization of the algebra as that of the EFT for gravitational observables at the background reference state $|R\rangle$. Note that the above construction inherently depends on the reference state. On the other hand, we may also consider other types of orthogonalization other than the Gram-Schmidt orthogonalization. We speculate that this ambiguity originates from field redefinitions in the EFT. It is very interesting to prove the above statements on EFT for gravitational observables. We leave this question for future investigations.

\subsubsection*{Gram-Schmidt orthogonalization for the ordered chord number states $|n\rangle~(0\leq n\leq L-1)$: Krylov basis}
We consider the Gram-Schmidt orthonormalization of the ordered chord number basis $|n\rangle~(0\leq n\leq L-1)$. We denote the resulting orthogonal basis as $|n^K\rangle~(0\leq n\leq L-1)$ and associated chord number operator as $\hat{N}^{K}$. This definition of length in JT gravity was considered in \cite{Miyaji:2024ity, Miyaji:2025yvm}. Remarkably, we can show that $\hat{N}^{K}$ in the ETH matrix model of the DSSYK and the Krylov spread complexity operator $\hat{N}^{\text{Krylov}}$ for initial state $|0\rangle$ are \emph{equivalent}. To see this, notice that the function $W_n(E)$ is a degree $n$ polynomial of $E$, whose coefficient of $E^n$ is non-zero. As the result, the finite $L$ chord number state can be expressed as
\begin{equation}\label{eq:showingequivalence}
    |n\rangle=W_n(H)|0\rangle=\sum_{i=0}^nc_n(i)H^i|0\rangle,~c_n(n)\neq0.
\end{equation}
This implies that the Gram-Schmidt orthogonalization of the ordered finite $L$ chord number states $|n\rangle,~(n=0,\cdots,L-1)$ is equivalent to that of the ordered states $H^n|0\rangle,~(n=0,\cdots,L-1)$, which is precisely the Krylov basis. Thus we can conclude
\begin{equation}
    \hat{N}^{K}=\hat{N}^{\text{Krylov}}.
\end{equation}
Here we note that the details of the wavefunction $W_n(E)$ do not affect the resulting operator $\hat{N}^K$ at all. Replacing $W_n(E)$ by any degree $n$ polynomial of $E$ does not modify the operator  $\hat{N}^K$. In particular, the finite $L$ chord number basis is not required to be nearly orthogonal with each other even at large $L$. Basically, $\hat{N}^{K}$ counts the polynomial degree of the wave function.

\subsubsection*{Gram-Schmidt orthogonalization from finite chord number state $|R\rangle$}
We consider a generalization of the above Gram-Schmidt orthogonalization to more general ordered chord number states. For $R\in\mathbb{Z}_{\geq 0}$, we consider the ordered $L$ states
\begin{equation}
    |R\rangle,~|R\pm1\rangle,~|R\pm2\rangle,~\cdots,
\end{equation}
with the understanding that when $R<(L-1)/2$, we will skip
$|n\rangle$ with negative $n$. We denote the orthogonalized basis as $|m^{M(R)}\rangle$, and corresponding finite $L$ chord number operator as $\hat{N}^{M(R)}$. As a special case, this operator at $R=0$ coincides with the Krylov spread complexity $\hat{N}^K$ with initial state $|0\rangle$ as
\begin{equation}
    \hat{N}^K=\hat{N}^{M(0)}.
\end{equation}
Note this identity is no longer true for general $R$, even if we take $|R\rangle$ as the initial state for the Krylov basis
\footnote{We note that we can consider another construction of chord number operator, by taking the ordered $L$ states
\begin{equation}
    |R\rangle,~|R+1\rangle,~|R+2\rangle,\cdots,
\end{equation}
and the orthogonalized basis $|m^{L(R)}\rangle$. The corresponding finite $L$ chord number operator $\hat{N}^{L(R)}$ is similar to the Krylov state complexity for the initial state $|R\rangle$, although it is in general distinct as we will see in \ref{section:Gaussian}. Note that $\hat{N}^{L(R)}$ cannot reproduce classical behavior for chord number $n<R$, thus we will not study $\hat{N}^{L(R)}$ further.}.

\subsubsection*{Other constructions of chord number operators}

We first explain the uniform orthogonalization. Let us choose the first $|m\rangle~(0\leq m\leq L-1)$ chord number states to be orthogonalized. Let us denote the restricted Gram matrix $(G_{nm}(H))_{0\leq n,m\leq L-1}$ as $G$, which is a real symmetric matrix. From this restricted Gram matrix, we can construct the following orthogonal basis
\begin{equation}
    |n^U\rangle:=\sum_{m=0}^{L-1} (G^{-1/2})_{mn}|m\rangle.
    \label{eq:nU}
\end{equation}
Note that this procedure is distinct from the Gram-Schmidt orthogonalization. We also note that $G_{nm}$ defined in \eqref{eq:G-def} depends on the random matrix $H$, which should not be confused with
the average of the inner product $\mathbb{E}[\bra n|m\ket]$. Using this orthonormal basis, we can define the finite $L$ chord number operator $\hat{N}^{U}$ by
\begin{equation}
    \hat{N}^{U}:=\sum_{n=0}^{L-1}n|n^U\rangle\langle n^U|.
    \label{eq:NU}
\end{equation}
By definition, $\hat{N}^{U}$ is a linear Hermitian operator whose eigenstates are $|n^U\rangle$ with eigenvalue $n^U$. 

Next, we consider the unorthogonalized finite chord number state, in which case we do not perform orthogonalization and use the non-orthogonal chord number state directly. Let us again consider $L$ chord number states $|n\rangle~(n\in\mathcal{S})$. We define the chord number operator via
\begin{equation}
    \hat{N}^{UN(R)}:=\sum_{n\in\mathcal{S}}n|n\rangle\langle n|.
    \label{eq:Unorthogonal}
\end{equation}
This new operator does not coincide with any of the operators defined above based on orthogonalization.

\subsection{Dynamics of the chord number}
We will consider the dynamics of the chord number of the time-evolved TFD state 
\begin{equation}
    |\text{TFD}(\beta,t)\rangle:=e^{-iHt}|\text{TFD}(\beta)\rangle.
\end{equation}
By definition, the initial state coincides with the chord number zero state $|0\rangle$. We will compute the expectation value of chord number operator $\hat{N}^X$
\begin{equation}
    \begin{split}
    \langle\text{TFD}(\beta,t)|\hat{N}^X|\text{TFD}(\beta,t)\rangle
    &=\sum_{i,j=1}^L\sum_{n^X}e^{i(E_i-E_j)t-\beta(E_i+E_j)/2}\langle E_i|n^X\rangle n^X\langle n^X|E_j\rangle.
    \end{split}
\end{equation}
For simplicity, we will set $\bt=0$ in what follows.
We will use the facts that both $|E_i\rangle$ and $|n^X\rangle$ are the complete orthonormal basis of the Hilbert space. Then the expectation value of $\hat{N}^X$ becomes
\begin{equation}
\begin{aligned}
\bra\hat{N}^X\ket &:=\frac{\langle\text{TFD}(t)|\hat{N}^X|\text{TFD}(t)\rangle}{
\langle\text{TFD}(t)|\text{TFD}(t)\rangle }\\
&=\frac{1}{L}\sum_{i,j=1}^L\sum_{n^X}e^{i(E_i-E_j)t}\langle E_i|n^X\rangle n^X\langle n^X|E_j\rangle.
\end{aligned}
\label{eq:N-ave}
\end{equation}

At late times, only the diagonal term $i=j$ in \eqref{eq:N-ave} survives due to the oscillating factor $e^{\ri t(E_i-E_j)}$. 
Thus we find
\begin{equation}
\begin{aligned}
\lim_{t\to\infty}\bra\hat{N}^X\ket &=\frac{1}{L}
\sum_{n^X}\sum_{i=1}^L\langle E_i|n^X\rangle n\langle n^X|E_i\rangle\\
&=\frac{1}{L}\sum_{n^X}n^X.
\end{aligned}
\label{eq:plateau}
\end{equation}
Note also that using \eqref{eq:0-sum}, one can see that $\bra\hat{N}^X\ket$ in \eqref{eq:N-ave} is written as
\begin{equation}
    \bra\hat{N}^X\ket=\sum_{n=0}^{L-1}\frac{1}{L}\bra 0|e^{\ri tH}|n^X\ket n\bra n^X|e^{-\ri tH}|0\ket.
    \label{eq:N-simple}
\end{equation}
In the next section, we will perform an explicit evaluation in the $q=0$ ETH matrix model, i.e. the Gaussian matrix model.

\section{Gaussian case $q=0$}\label{section:Gaussian}

In this section, we consider the ETH matrix model of DSSYK at $q=0$, where the potential becomes Gaussian and many expressions can be obtained analytically. After explaining these simplifications, we will study the dynamics of the finite $N$ chord number.

\subsection{Finite $N$ chord number state at $q=0$}

\begin{figure}[tb]
\centering
\includegraphics[width=0.5\linewidth]{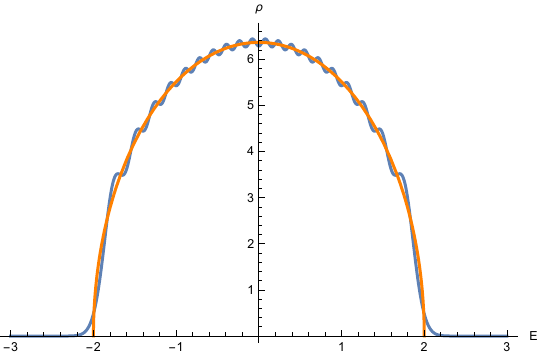}
  \caption{
Plot 
of $\rho(E)$ (blue) and $L\rho_0(E)$ (orange) for $L=20$
}
  \label{fig:gaussian-density}
\end{figure}

Let us consider the ETH matrix model of DSSYK in \eqref{eq:ETH} at $q=0$, whose potential becomes Gaussian 
\begin{equation}
\begin{aligned}
 \mathcal{Z}(\beta)=\int_{L\times L} dH e^{-\frac{L}{2}\Tr H^2}\Tr[e^{-\beta H}] .
\end{aligned} 
\label{eq:ETH-Gauss}
\end{equation}
Therefore we have
\begin{equation}
\begin{aligned}
 \mathcal{Z}(\beta)=\int\prod_{i=1}^LdE_ie^{-\frac{L}{2}E_i^2}\prod_{i<j}(E_i-E_j)^2\sum_{i=1}^Le^{-\beta E_i}.
\end{aligned} 
\end{equation}
When $q=0$, the wavefunction $\bra n|\th\ket$ reduces to the
Chebyshev polynomial of the second kind $U_n(E/2)$ with $E=2\cos\th$
\begin{equation}
\begin{aligned}
 \bra n|\th\ket=\sum_{k=0}^ne^{\ri(n-2k)\th}=\frac{\sin(n+1)\th}{\sin\th}=U_n(E/2).
\end{aligned} 
\end{equation}
The measure factor is $\mu(\th)=(q,e^{\pm2\ri\th};q)_\infty\big|_{q=0}=4\sin^2\th$, 
which agrees with the Wigner 
semicircle distribution of the Gaussian matrix model
\begin{equation}
\begin{aligned}
 \frac{d\th}{2\pi}\mu(\th)=dE\rho_0(E),\qquad \rho_0(E)=\frac{1}{2\pi}\rt{4-E^2}.
\end{aligned}
\label{eq:rho0-Wigner} 
\end{equation}
The $n$-th orthogonal polynomial $P_n(E)$ is given by the Hermite polynomial
\begin{equation}
\begin{aligned}
 P_n(E)=(2L)^{-\frac{n}{2}}H_n(x),\quad x=\rt{\frac{L}{2}}E,\quad
h_n=\rt{2\pi} n! L^{-n-\hf}.
\end{aligned} 
\end{equation}
After some algebra, one can show that $\rho(E)$ is given by
\begin{equation}
\begin{aligned}
 \rho(E)=\frac{e^{-\frac{L}{2}E^2}}{\rt{2\pi}2^{L}(L-1)!L^{-\hf}}
\Bigl[H_L(x)^2-H_{L+1}(x)H_{L-1}(x)\Bigr].
\end{aligned} 
\end{equation}
In the large $N$ limit, $\rho(E)$ is expanded as
\begin{equation}
\begin{aligned}
 \rho(E)=\sum_{g=0}^\infty L^{1-2g}\rho_g(E).
\end{aligned} 
\end{equation}
As we can see from Figure~\ref{fig:gaussian-density},
$\rho(E)$ is approximated by $L\rho_0(E)$ for large $N$.

\subsection{Gram matrix is non-diagonal}

\begin{figure}[h]
    \centering
\subcaptionbox{$k=5$\label{sfig:i-k5}}{
\includegraphics[keepaspectratio,scale=0.6,width=0.3\linewidth]{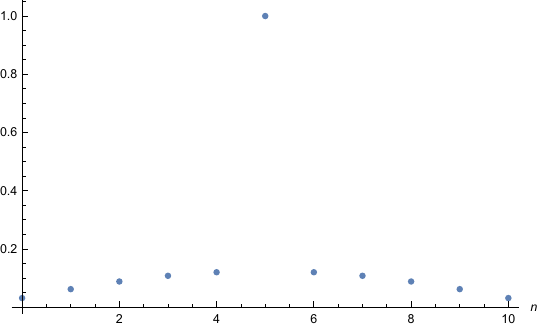}}
\hskip2mm
\subcaptionbox{$k=10$\label{sfig:i-k10}}{
\includegraphics[keepaspectratio,scale=0.6,width=0.3\linewidth]{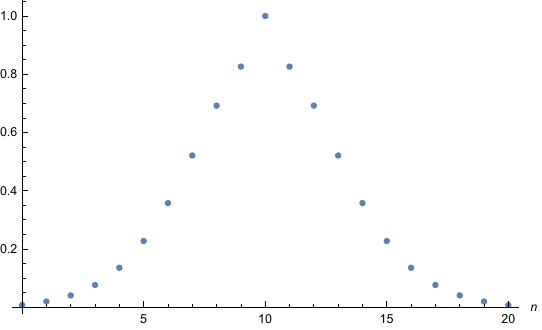}}
\hskip2mm
\subcaptionbox{$k=15$\label{sfig:i-k15}}{
\includegraphics[keepaspectratio,scale=0.6,width=0.3\linewidth]{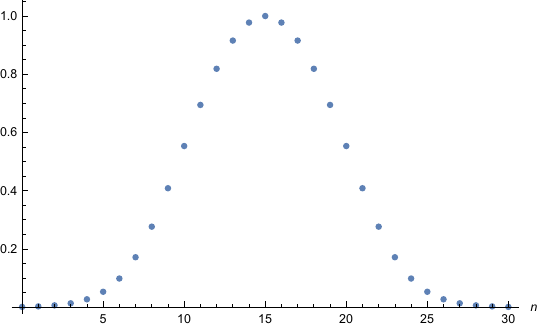}}
\caption{Plot of the normalized inner product $(n,2k-n)$
for $n=0,\cdots,2k$ for $L=20$.}
\label{fig:inner}
\end{figure}

Let us consider the expectation value of the 
inner product $\bra n|m\ket$ \eqref{eq:modified-inner} at finite $L$.
In Figure~\ref{fig:inner}, we show the plot of normalized inner product 
\begin{equation}
    (n,2k-n):=\frac{\mathbb{E}[\bra n|2k-n\ket]}{\rt{\mathbb{E}[\bra n|n\ket]\mathbb{E}[\bra 2k-n|2k-n\ket]}}
\end{equation} 
for $n=0,\cdots,2k$, where the expectation value $\mathbb{E}$ is defined by the average in the $q=0$
ETH matrix model, i.e.
the Gaussian matrix model
\begin{equation}
\mathbb{E}\bigl[f(H)\bigr]=\frac{\int dH e^{-\frac{L}{2}\Tr H^2}f(H)}{\int dH e^{-\frac{L}{2}\Tr H^2}}.
\label{eq:ETH-Gauss}
\end{equation}
When the 
inner product $\bra n|m\ket$ is diagonal, $(n,2k-n)$ is localized to 
$n=k$. As we can see from Figure~\ref{fig:inner}, 
this is the case for small $k$.
However, as $k$ increases, the non-diagonal part of 
$(n,2k-n)$ becomes non-negligible.
Note that
\begin{equation}
\begin{aligned}
   (n,2k+1-n)= 0,\quad (0\leq n\leq 2k+1),
\end{aligned} 
\end{equation}
since $\rho(E)$ is an even function of $E$.

Let us verify the above claim analytically. In terms of $x=\rt{L/2}E$, we define $\h{\rho}(x)$ by
\begin{equation}
\begin{aligned}
 dE\rho(E)=dx\h{\rho}(x).
\end{aligned} 
\end{equation}
We find
\begin{equation}
\begin{aligned}
 \h{\rho}(x)=L\psi_L(x)^2-\rt{L(L+1)}\psi_{L+1}(x)\psi_{L-1}(x)
\end{aligned} 
\end{equation}
where $\psi_n(x)$ is defined by
\begin{equation}
\begin{aligned}
 \psi_n(x)=\frac{e^{-\hf x^2}}{(\rt{\pi}2^n n!)^{\hf}}H_n(x),\quad
\int_{-\infty}^\infty dx\psi_n(x)\psi_m(x)=\cob_{n,m}.
\end{aligned} 
\end{equation}
$\psi_j(x)$ is the wavefunction of harmonic oscillator $[a,a^\dagger]=1$
\begin{equation}
\begin{aligned}
 \psi_j(x)=\bra x|j\ket_a,\quad |j\ket_a=\frac{(a^\dag)^j}{\rt{j!}}|0\ket_a.
\end{aligned} 
\end{equation}
Here we have put the subscript $a$ in $|j\ket_a$ in order to distinguish it
from the length state $|n\ket$. 
Then we can compute the one-point function of $f(E)$ from
the matrix element of the harmonic oscillator
\begin{equation}
\begin{aligned}
 \int dE\rho(E)f(E)&=L\cdot{}_a\bra L|f\Bigl(\frac{a+a^\dag}{\rt{L}}\Bigr)|L\ket_a
-\rt{L(L+1)}\cdot{}_a\bra L+1|f\Bigl(\frac{a+a^\dag}{\rt{L}}\Bigr)|L-1\ket_a.
\end{aligned} 
\end{equation}
Using this method, we find the expectation value of the inner product $\mathbb{E}\bigl[\bra n|m\ket\bigr]$ for small $n+m$
\begin{equation}
\begin{aligned}
 \bra 1|1\ket&=1,\quad \bra0|2\ket=0,\\
\bra2|2\ket&=1,\quad \bra1|3\ket=\frac{1}{L^2},\quad
\bra0|4\ket=\frac{1}{L^2},\\
\bra3|3\ket&=1+\frac{6}{L^2},\quad
\bra2|4\ket=\frac{6}{L^2},\quad
\bra1|5\ket=\frac{6}{L^2},\quad
\bra0|6\ket=\frac{5}{L^2},\\
\bra4|4\ket&=1+\frac{21}{L^2}+\frac{21}{L^4},\quad 
\bra3|5\ket=\frac{21}{L^2}+\frac{21}{L^4},\quad \cdots.
\end{aligned} 
\end{equation}
Here we have suppressed the symbol $\mathbb{E}$ of expectation value for brevity.
More generally, we find the exact form of $\bra0|2k\ket$
\begin{equation}
\begin{aligned}
 \bra0|2k\ket=(L-1)!\sum_{i=0}^{k}\sum_{j=0}^{k-i}\frac{(-1)^i (2k-i)!(2L)^{i-k}2^j}{i!j!(j+1)!(L-1-j)!(k-i-j)!}.
\end{aligned} 
\label{eq:02k-exact}
\end{equation}
We observe that
\begin{equation}
\bra 0|2k\ket=\left\{
\begin{aligned}
 &1,\quad &&(k=0),\\
&\binom{k+2}{4}\frac{1}{L^2}+\cO(L^{-4}),\quad
&&(k\geq1).
\end{aligned} \right.
\label{eq:02k-genus}
\end{equation}
Using the relation
\begin{equation}
\begin{aligned}
 U_n(x)U_m(x)=\sum_{j=0}^{\min(n,m)}U_{n+m-2j}(x),
\end{aligned} 
\end{equation}
the inner product $\bra n|m\ket$ is written as a combination of
$\bra 0|2k\ket$ in \eqref{eq:02k-exact}
\begin{equation}
\begin{aligned}
 \bra n|m\ket=\sum_{j=0}^{\min(n,m)}\bra 0|n+m-2j\ket.
\end{aligned} 
\end{equation}
Thus, we find that $\bra n|m\ket=0$ for odd $n+m$, and for even $n+m$ with $n+m=\mathcal{O}(L^0)$, we conclude
\begin{equation}
\bra n|m\ket=\delta_{nm}+L^{-2}R_{nm},\
\end{equation}
where $R_{nm}=\mathcal{O}(L^0)$. This relation is analogous to \cite{Penington:2019kki} and the ETH. It would be interesting to understand whether this behavior persists for $n,~m\in\mathcal{L^1}$. 


\subsubsection*{Comparison with discrete Weil-Petersson volume}
We explain the discrete Weil-Petersson volume in the Gaussian case. The higher genus corrections to the inner product $\bra 0|2k\ket$ is given by
\begin{equation}
\begin{aligned}
 \sum_{g=1}^\infty L^{-2g}\sum_{b=1}^\infty b V_{g,1}(b) \bra 0|\cB_b|2k\ket
\end{aligned} 
\end{equation}
where $V_{g,n}(\bold{b})$ is the discrete volume of the moduli space of 
Riemann surface of genus-$g$ with $n$-boundaries.
For the Gaussian matrix model, $V_{g,1}(b)$ is given by
\cite{norbury2013polynomials,Okuyama:2023kdo}
\begin{equation}
\begin{aligned}
 V_{1,1}(b)&=\frac{b^2-2^2}{48}P_b,\\
V_{2,1}(b)&=\frac{(b^2-2^2)(b^2-4^2)(b^2-6^2)(5b^2-32)}{8847360}P_b,
\end{aligned} 
\end{equation}
where $P_b=\frac{1+(-1)^b}{2}$ is the projection to even $b$.
The amplitude $\bra 0|\cB_b|2k\ket$
of baby universe operator $\cB_b$ is given by
\begin{equation}
\begin{aligned}
 \bra 0|\cB_b|2k\ket=\int_0^\pi\frac{d\th}{2\pi}\,2\cos(b\th)U_{2k}(\cos\th)
=\left\{
\begin{aligned}
&1,\quad&&(0\leq b\leq 2k,~b=\text{even}),\\
&0,\quad &&(\text{otherwise}).
\end{aligned}
\right.
\end{aligned} 
\label{eq:baby-q0}
\end{equation}
Then the genus-one correction becomes
\begin{equation}
\begin{aligned}
 \sum_{b=1}^{2k}bV_{1,1}(b)=\frac{(k-1)k(k+1)(k+2)}{24}=\binom{k+2}{4},
\end{aligned} 
\end{equation}
which reproduces the result in \eqref{eq:02k-genus}.
For genus-two, we find
\begin{equation}
\begin{aligned}
 \sum_{b=1}^{2k}bV_{2,1}(b)=\frac{
(k-3) (k-2) (k-1)^2 k (k+1) (k+2)^2 (k+3) (k+4)}{34560}.
\end{aligned} 
\end{equation}
One can check that this is consistent with the exact result \eqref{eq:02k-exact}.

\subsection{Finite $L$ chord number basis}

When $q=0$, we have
\begin{equation}
    \langle E_i|n\rangle=U_n(E_i/2)=\frac{\sin(n+1)\theta_i}{\sin\theta_i},~E_i=2\cos\theta_i.
\end{equation}
In the following, we will explicitly write the first few $|n^X\rangle$ for various basis, $X=K,~M(R),~L(R),~U$, as well as the Krylov basis with the initial state $|R\rangle$, denoted as $X=K(R)$.


\subsubsection*{$X=K$: Krylov basis}
The Krylov basis states for small $n$ are given by
\begin{equation}
\begin{split}
    |0^K\rangle&=\frac{1}{\sqrt{L}}|0\rangle=\frac{1}{\sqrt{L}}\sum_{i=0}^{L-1}|E_i\rangle,\\
    |1^K\rangle&\propto\sum_{i=0}^{L-1}(E_i-\sum_-\frac{1}{L}E_i)|E_i\rangle.
    \end{split}
\end{equation}

\subsubsection*{$X=K(R)$: Krylov basis with initial state $|R\rangle$}
We consider the Krylov basis with the initial state $|R\rangle$. We have
\begin{equation}
\begin{split}
    |R^{K(R)}\rangle&=\frac{|R\rangle}{\sqrt{\langle R|R\rangle}}=\frac{U_R(H/2)|0\rangle}{\sqrt{\langle R|R\rangle}},
    \\
    |R+1^{K(R)}\rangle&\propto \Big(HU_R(H/2)-\frac{\langle R|H|R\rangle}{\langle R|R\rangle}U_R(H/2)\Big)|0\rangle.
    \end{split}
\end{equation}


\subsubsection*{$X=M(R),~L(R)$: Chord number basis around $|R\rangle$}
For $A=M,~R$, we have
\begin{equation}
\begin{split}
    |R^{A(R)}\rangle&=\frac{|R\rangle}{\sqrt{\langle R|R\rangle}}=\frac{U_R(H/2)|0\rangle}{\sqrt{\langle R|R\rangle}},
    \\
    |R+1^{A(R)}\rangle&\propto \Big(U_{R+1}(H/2)-\frac{\langle 0|U_{R+1}(H/2)U_R(H/2)|0\rangle}{\langle R|R\rangle}U_R(H/2)\Big)|0\rangle.
    \end{split}
\end{equation}
From this expression, we can confirm that this basis is distinct from the Krylov basis $X=K(R)$, due to the contribution $\langle 0|U_{R-1}(H/2)U_R(H/2)|0\rangle$, which is subleading in $1/L$. However, once we proceed and consider $|R-1^{M(R)}\rangle$, there is no longer a direct connection to the usual Krylov state complexity.


\subsubsection*{$X=U$: Uniform orthogonalization}

In order to find the basis \eqref{eq:nU} in the uniform orthogonalization, 
we need to know $G^{-1/2}$ which is not easy to compute in general. 
However, the length operator
$\hat{N}_U$ in \eqref{eq:NU} is written as
\begin{equation}
\hat{N}_U=\sum_{m,l=0}^{L-1}|m\ket M_{ml}\bra l|
\label{eq:NU-M}
\end{equation}
with
\begin{equation}
    M_{ml}=\sum_{n=0}^{L-1}(G^{-\hf})_{mn}n(G^{-\hf})_{nl}.
\end{equation}
It turns out that $M_{ml}$ can be evaluated in a closed form for $L=2$
\begin{equation}
    \begin{aligned}
        M_{00}&=\frac{(E_1+E_2)^2}{(E_1-E_2)^2\bigl[(E_1+\vep)^2+(E_2-\vep)^2\bigr]},\\
        M_{01}=M_{10}&=-\frac{(E_1+E_2)(2+|E_1-E_2|)}{(E_1-E_2)^2\bigl[(E_1+\vep)^2+(E_2-\vep)^2\bigr]},\\
        M_{11}&=\frac{(2+|E_1-E_2|)^2}{(E_1-E_2)^2\bigl[(E_1+\vep)^2+(E_2-\vep)^2\bigr]},
    \end{aligned}
    \label{eq:M2}
\end{equation}
where $\vep$ denotes the sign of $E_1-E_2$
\begin{equation}
\begin{aligned}
 \vep=\text{sign}(E_1-E_2)=\frac{E_1-E_2}{|E_1-E_2|}.
\end{aligned} 
\end{equation}

\subsection{Chord number dynamics at late time}
In this section, we will study the dynamics of the chord number expectation value
\eqref{eq:N-simple}.
We take the ensemble average of $\{E_i\} $
with respect to the measure of the ETH matrix model for $q=0$ in \eqref{eq:ETH-Gauss}
\begin{equation}
    \bra \hat{N}^X\ket=\mathbb{E}\Biggl[\frac{1}{L}\bra0|e^{\ri tH} \hat{N}^Xe^{-\ri tH}|0\ket\Biggr].
    \label{eq:hatN-E}
\end{equation}
Again, we should emphasize that the expectation value $\mathbb{E}$ in \eqref{eq:hatN-E} is taken after we perform the orthogonalization
$|n\ket\to|n^X\ket$ in \eqref{eq:N-simple} for a fixed instance of the random matrix $H$. Below we compute $ \bra \hat{N}^X\ket$  at small $L$ for several definition of the length operator
$\hat{N}^X$.

\subsubsection*{Uniform orthogonalization $\bra \hat{N}^U\ket$}

\begin{figure}[tb]
\centering
\subcaptionbox{$L=2$\label{sfig:N2-Mhalf}}{\includegraphics[width=0.4\linewidth]{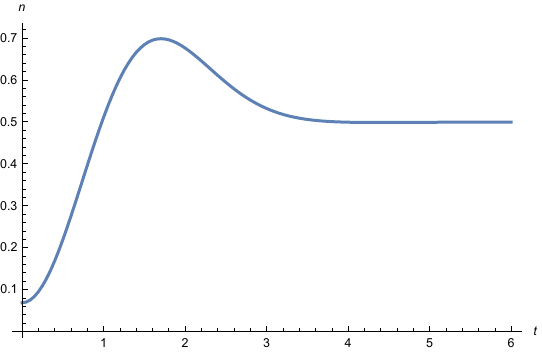}}
\hskip8mm
\subcaptionbox{$L=3$\label{sfig:N3-Mhalf}}{\includegraphics[width=0.4\linewidth]{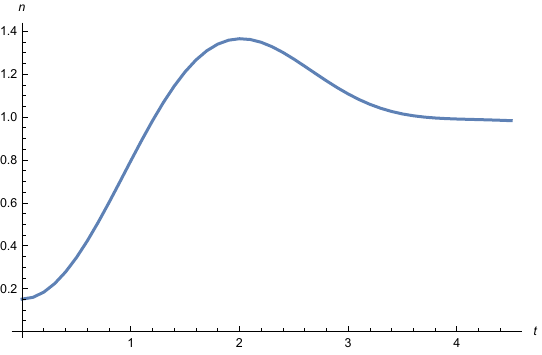}}
\caption{Plot 
of $\bra \hat{N}^U\ket$  for \subref{sfig:N2-Mhalf} $L=2$ and \subref{sfig:N3-Mhalf} $L=3$. In both cases, we see the ramp-slope-plateau behavior and the saturation at the value $(L-1)/2$. In Mathematica, $G^{-\hf}$ can be computed by the command $G^{-\hf}=\texttt{MatrixPower}[G,-1/2]$. In this way, we have numerically computed $\bra \hat{N}^U\ket$ for $L=3$.
}
 \label{fig:NU}
\end{figure}

From \eqref{eq:NU-M} and \eqref{eq:M2}, $\bra\hat{N}^U\ket$ for $L=2$ is given by
\begin{equation}
\begin{aligned}
 \bra\hat{N}^U\ket=&\hf+\frac{1}{\pi}\int_{-\infty}^\infty\prod_{i=1,2}dE_ie^{-E_i^2}
\frac{(E_1-E_2)^2(E_1+\vep)(E_2-\vep)}{(E_1+\vep)^2+(E_2-\vep)^2}\cos \bigl[(E_1-E_2)t\bigr]
\end{aligned}
\label{eq:n-Mhalf} 
\end{equation}
where $\vep$ denotes
the sign of $E_1-E_2$.
In Figure~\ref{sfig:N2-Mhalf}, we show the plot of 
$\bra \hat{N}^U\ket$ for $L=2$ in \eqref{eq:n-Mhalf}.
We can see that $\bra \hat{N}^U\ket$ increases linearly at early times and 
then decreases, and finally plateaus at late times.
This is consistent with the ``ramp-slope-plateau'' behavior of the spread complexity. However, note that $\bra \hat{N}^U\ket$ is nonzero at the initial state. This is due to the uniform orthogonalization, which perturbs $|0^U\rangle$ away from $|0\rangle$.
We can see that $\bra\hat{N}^U\ket$ approaches $\frac{L-1}{2}$ at late times, as expected.

\subsubsection*{Krylov complexity $\bra \hat{N}^K\ket$}

For $L=2$, the Krylov state complexity $\bra \hat{N}^K\ket$ is given by
\begin{equation}
\begin{aligned}
 \bra\hat{N}^K\ket
&=\frac{1}{\pi}
\int dE_1dE_2 e^{-(E_1^2+E_2^2)}(E_1-E_2)^2\sin^2\Biggl[\frac{t}{2}(E_1-E_2)\Biggr]\\
&=\hf +\hf (t^2-1)e^{-\frac{t^2}{2}},
\end{aligned} 
\end{equation}
as was obtained in \cite{Caputa:2024vrn}.
\begin{figure}[tb]
\centering
\subcaptionbox{$\bra\hat{N}^K\ket$ for $L=2$\label{sfig:L2}}{\includegraphics[width=0.4\linewidth]{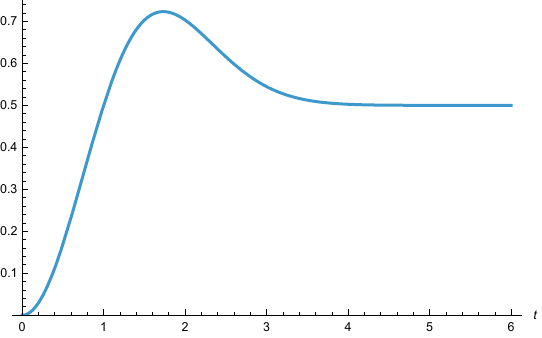}}
\hskip8mm
\subcaptionbox{$\bra\hat{N}^{UN(0)}\ket$ for $L=2$\label{sfig:Un}}{\includegraphics[width=0.4\linewidth]{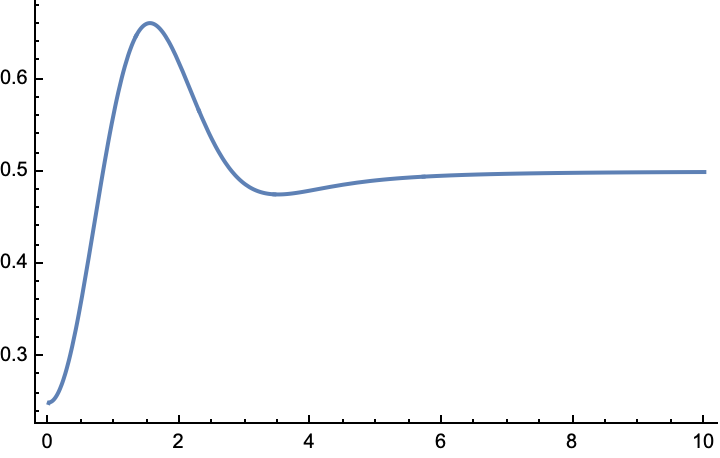}}
\caption{Plot 
of \subref{sfig:L2} $\bra \hat{N}^U\ket$ and \subref{sfig:Un} $\bra\hat{N}^{UN(0)}\ket$ for $L=2$. In both cases, we see the ramp-slope-plateau behavior and the saturation at the value $(L-1)/2$.
}
 \label{fig:L2-GS}
\end{figure}
As we can see from Figure~\ref{fig:L2-GS}, $\bra\hat{N}^K\ket$ vanishes at $t=0$ and approaches $(L-1)/2$ at late times, displaying the ramp-slope-plateau behavior. In the current case $L=2$, $\bra \hat{N}^K\ket$ is precisely identical to the probability distribution $P(n^K)$ at $n^K=1$. In contrast to $\hat{N}^U$, we have $\bra\hat{N}^K\ket_{t=0}=0$ for general $L$, because $|0^K\ket$ is not perturbed from $|0\ket$.

The "ramp-slope-plateau" structure in Krylov spread complexity is confirmed in more general $L$ and $q$ in DSSYK. Indeed, using the random matrix technique, the plateau \cite{Nandy:2024zcd} and the slope \cite{Balasubramanian:2024lqk} were also verified for the ETH matrix model for DSSYK model for general $L$ and $q$.

\subsubsection*{Un-orthogonalized $\bra \hat{N}^{UN(0)}\ket$}
For $L=2$, $\bra \hat{N}^{UN(0)}\ket$ is given by
\begin{equation}
\begin{aligned}
\hat{N}^{UN(0)}
&=\frac{1}{\pi}
\int dE_1dE_2 e^{-(E_1^2+E_2^2)}(E_1-E_2)^2\Biggl(\frac{1}{2}+\frac{E_1E_2}{E_1^2+E_2^2}\cos\Biggl[t(E_1-E_2)\Biggr]\Biggr)\\
&=\frac{t^4-12+e^{-\frac{t^2}{2}}\Big(t^6+t^4+6t^2+12\Big)}{2t^4}.
\end{aligned} 
\end{equation}
As can be seen from Figure~\ref{fig:L2-GS}, the qualitative behavior is similar to $\hat{N}^U$; it is nonzero at $t=0$ but still displays the ramp-slope-plateau behavior, saturating at the value $\frac{L-1}{2}$.

\subsubsection*{Finite $R$ orthogonalization $\bra \hat{N}^{L(R)}\ket$, $\bra \hat{N}^{M(R)}\ket$ and $\bra \hat{N}^{K(R)}\ket$}
The chord number operators with reference state $|R\rangle$ are degenerate when $L=2$, namely we have $\bra\hat{N}^{L(R)}\ket=\bra \hat{N}^{M(R)}\ket=\bra \hat{N}^{K(R)}\ket$. When $R=1$, the expectation value is given by
\begin{equation}
\begin{aligned}
 \bra\hat{N}^{M(1)}\ket
&=\frac{1}{\pi}
\int dE_1dE_2 e^{-(E_1^2+E_2^2)}(E_1-E_2)^2\\
&\times\Big(1\cdot\Big(\frac{1}{2}+\frac{E_1E_2}{E_1^2+E_2^2}\cos\Biggl[t(E_1-E_2)\Biggr]\Big)+2\cdot\Big(\frac{1}{2}-\frac{E_1E_2}{E_1^2+E_2^2}\cos\Biggl[t(E_1-E_2)\Biggr]\Big)\Big)\\
&=\frac{3(t^4+4)-e^{-\frac{t^2}{2}}\Big(t^6+t^4+6t^2+12\Big)}{2t^4}.
\end{aligned} 
\end{equation}
\begin{figure}[tb]
\centering
\subcaptionbox{Chord number expectation value $\bra \hat{N}^{M(1)}\ket$\label{sfig:N(1)}}{\includegraphics[width=0.4\linewidth]{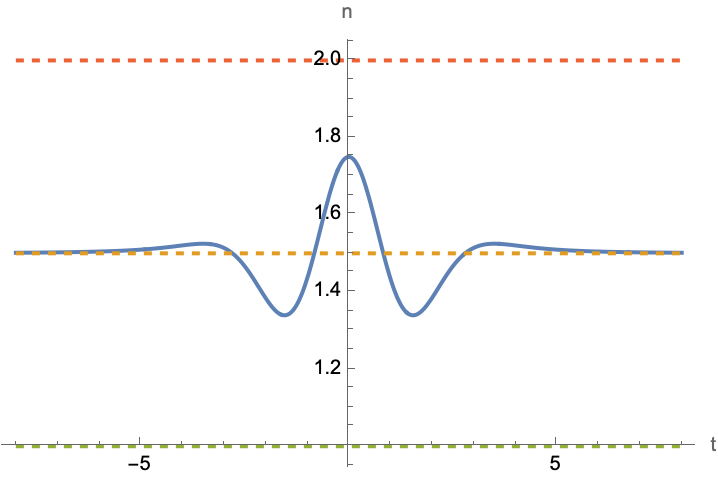}}
\hskip8mm
\subcaptionbox{Chord number probability distribution $P(n^{M(1)})$ \label{sfig:N3-Mhalf}}{\includegraphics[width=0.4\linewidth]{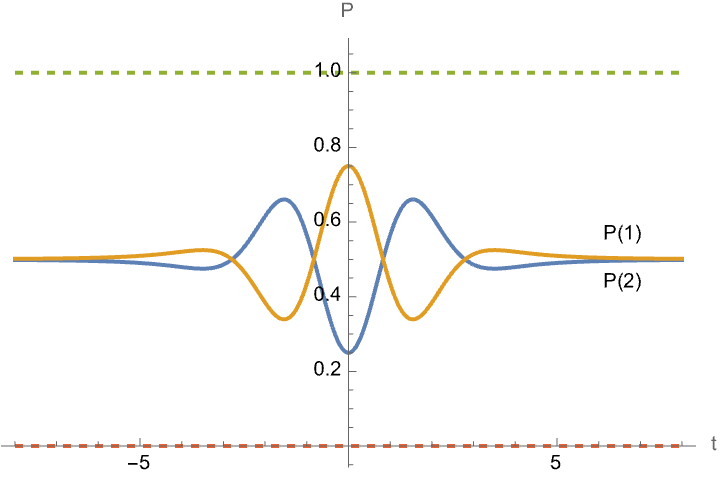}}
\caption{Plot of the expectation value $\bra\hat{N}^{L(R)}\ket=\bra \hat{N}^{M(R)}\ket=\bra \hat{N}^{K(R)}\ket$ for $L=2,~R=1$ and the chord number probability distribution $P(n^{M(1)})$. }
 \label{fig:NR}
\end{figure}
The chord number probability distribution $P(n^X)$ is
\begin{equation}
\begin{aligned}
 P(n^X)&=\frac{t^4+(-1)^{n^X}12+(-1)^{n^X+1}e^{-\frac{t^2}{2}}\Big(t^6+t^4+6t^2+12\Big)}{2t^4},~(n^X=1,~2).
\end{aligned} 
\end{equation}
As we can see from Figure~\ref{fig:NR}, the qualitative behavior is radically different from other chord number operators. We can see from the peak of $P(2)$ in Figure (b), $n^X=1$ becomes dominant at finite $t$. At large $L$ such a peak is expected to become sharp, giving the concentration of probability such that classical geometry is realized. Such phenomenon was considered in JT gravity in \cite{Miyaji:2024ity}.


\section{Discussion}\label{section:discussion}


In this paper, we studied constructions of chord number operators in each instance of ensemble in the ETH matrix model for double-scaled SYK. Our strategy is to orthogonalize a subset of the chord number states, which can span the entire Hilbert space and on which the action of $q$-oscillator is realized approximately. We interpret this dependence on the reference state as closely related to that of the EFT, which can describe perturbation around the reference state, as a realization of ``non-isometric code'' of \cite{Akers:2022qdl} applied to gravitational observables. To establish such interpretation on a firmer footing, it would be interesting to investigate the gravitational description of such generalized chord number operators in terms of the sine-dilaton gravity \cite{Blommaert:2024whf, Blommaert:2025avl}. In the present article, we have concentrated on the theory without matter. To relate the current discussion with the firewall, it is necessary to include matter in the bulk, which requires the introduction of matter chords \cite{Lin:2022rbf, Lin:2023trc, Xu:2024hoc, Xu:2024gfm, Ambrosini:2024sre}. By the introduction of matter chords, we should be able to construct a new basis which accommodates approximate representation of the algebra of DSSYK observables. We have also only considered the chord number operator, but the analysis can be extended to wormhole velocity \cite{Iliesiu:2024cnh} and time-shift \cite{Miyaji:2024ity}. As we commented in the article, the set of rules for chord diagrams is still unknown, which can reproduce the correlators of the ETH matrix model. It would be interesting to search for such a set of rules, along with the investigation of the geometry of the discrete baby universes and discrete Weil-Peterson volume \cite{Okuyama:2023kdo}.

\subsection*{Poincare recurrence}

The chord number states $|n\rangle~(n\in\mathbb{Z}_{\geq0})$ before the orthogonalization is expected to be dense in the non-perturbative Hilbert space. More presicely, for any $n\in\mathbb{Z}_{\geq0}$ and $\epsilon>0$, there exists $r(n:\epsilon)=\mathcal{O}(e^{e^{S_0}})$ such that $|\langle n|n+r(n:\epsilon)\rangle|\geq 1-\epsilon$. Thus arbitrarily similar states can have completely different interpretations. Furthermore, the length operators defined from these reference states can be completely different, as the actions of $q$-oscillators on these reference states can be drastically different. It would be very interesting to examine this phenomenon explicitly and to construct EFT for such a doubly exponentially large chord number state.


\acknowledgments
 We thank Masaki Shigemiori for early-stage collaborations. We are grateful to Chris Akers, Vijay Balasubramanian, Pawel Caputa, Pratik Nandy, Shanming Ruan, Tadashi Takayanagi, and Jiuci Xu for discussions. MM is supported in part by JSPS KAKENHI Grant Numbers 24K17044 and 25K00997. SM is supported by JST SPRING, Grant Number JPMJSP2125 and the “THERS Make New Standards Program for the Next Generation Researchers.” KO is supported in part by JSPS Grant-in-Aid for Transformative Research Areas (A) ``Extreme Universe'' 21H05187, and JSPS KAKENHI 22K03594 and 25K07300.
 


\bibliography{paper}
\bibliographystyle{utphys}

\end{document}